\documentclass[12pt]{article}
\usepackage{amsmath}
\usepackage{amsfonts}
\usepackage{graphicx,psfrag,epsf}
\usepackage{enumerate}
\usepackage{natbib}
\usepackage{url} 
\usepackage{bm}
\usepackage{color}

\newcommand{\blind}{0}

\begin{document}

\def\spacingset#1{\renewcommand{\baselinestretch}%
{#1}\small\normalsize} \spacingset{1}
\makeatletter
\newcommand{\figcaption}[1]{\def\@captype{figure}\caption{#1}}
\newcommand{\tblcaption}[1]{\def\@captype{table}\caption{#1}}
\makeatother





\if0\blind
{
  \title{\bf Inference for log Gaussian Cox processes using an approximate marginal posterior}
  \author{Shinichiro Shirota\thanks{Department of Statistical Science, Duke University, U.S. (E-mail: ss571@stat.duke.edu)}\hspace{.2cm}\\
    Department of Statistical Science, Duke University, U.S. \\
    and \\
    Alan. E. Gelfand\thanks{Department of Statistical Science, Duke University, U.S. (E-mail: alan@stat.duke.edu)} \\
    Department of Statistical Science, Duke University, U.S.}
  \maketitle
} \fi

\if1\blind
{
  \bigskip
  \bigskip
  \bigskip
  \begin{center}
    {\LARGE\bf Inference for log Gaussian Cox processes using an approximate marginal posterior}
\end{center}
  \medskip
} \fi

\bigskip
\begin{abstract}
The log Gaussian Cox process is a flexible class of point pattern models for capturing spatial and spatio-temporal dependence for point patterns.
Model fitting requires approximation of stochastic integrals which is implemented through discretization of the domain of interest.
With fine scale discretization, inference based on Markov chain Monte Carlo is computationally heavy because of the cost of repeated iteration or inversion or Cholesky decomposition (cubic order) of high dimensional covariance matrices associated with latent Gaussian variables.
Furthermore, hyperparameters for latent Gaussian variables have strong dependence with sampled latent Gaussian variables.
Altogether, standard Markov chain Monte Carlo strategies are inefficient and not well behaved.

In this paper, we propose an efficient computational strategy for fitting and inferring with spatial log Gaussian Cox processes.
The proposed algorithm is based on a pseudo-marginal Markov chain Monte Carlo approach.
We estimate an approximate marginal posterior for parameters of log Gaussian Cox processes and propose comprehensive model inference strategy.
We provide details for all of the above along with some simulation investigation for the univariate and multivariate settings.
As an example, we present an analysis of a point pattern of locations of three tree species, exhibiting positive and negative interaction between different species.
\end{abstract}

\noindent%
{\it Keywords:}  kernel mixture marginalization; Laplace approximation; multivariate Poisson log normal; pseudo marginal MCMC
\vfill

\newpage
\spacingset{1.45} 

\section{Introduction}
There is increasing collection of space and space time point pattern data in settings including point patterns of locations of tree species (\cite{Burslemetal(01)}, \cite{Wiegandetal(09)} and \cite{Illianetal(08)}), of locations of disease occurrences (\cite{Liangetal(09)}, \cite{RuizMorenoetal(10)} and \cite{Diggleetal(13)}), of locations of earthquakes (\cite{Ogata(99)} and \cite{MarsanLengline(08)}) and of locations of crime events (\cite{ChaineyRatcliffe(05)} and \cite{GrubesicMack(08)}).
In addition, the points may be observed over time (\cite{GrubesicMack(08)} and \cite{Diggleetal(13)}).
In the literature, the most commonly adopted class of models for such data are nonhomogeneous Poisson processes (NHPP) or, more generally log Gaussian Cox processes (LGCP) (see \cite{MollerWaagepetersen(04)} and references therein).
The intensity surface of a Cox process is a realization of a stochastic process.  However, given the intensity surface, Cox processes are Poisson processes.
The LGCP was originally proposed by \cite{Molleretal(98)} and extended to space-time case by \cite{BrixDiggle(01)}.
As the name suggests, the intensity function of this process is driven by the exponential of Gaussian processes (GP).

Fitting LGCP models is challenging because the likelihood of the LGCP includes an integration of the intensity over the domain of interest.  The integral is stochastic and is analytically intractable so approximation methods are required.
One strategy is to grid the study region (creating a so-called set of representative points) and approximate this integral with a Riemann sum (\cite{Molleretal(98)} and \cite{MollerWaagepetersen(04)}).
Bayesian model fitting using Markov chain Monte Carlo (MCMC) methods is more demanding since it requires repeated approximation over the iterations of the sampling.
As a result, a standard MCMC scheme requires repeated conditional sampling of high dimensional latent Gaussian variables.
The convergence of posterior samples based on this approximated likelihood to the exact posterior distribution is guaranteed by results in \cite{Waagepetersen(04)}.
However, sampling of high dimensional GP remains a demanding computational task.
Calculating the inverse or Cholesky decomposition is required for sampling of the joint distribution of $n$ variables.
These calculations for an $n$-dimensional covariance matrix require $\mathcal{O}(n^3)$ computational time and $\mathcal{O}(n^2)$ memory for storage.
Recent approaches for dealing with this "big n" problem in geostatistical contexts include the nearest-neighbor GP (\cite{Dattaetal(16a)}) and multi-resolution GP (\cite{Katzfuss(16)}).


As for the literature on MCMC-based Bayesian inference for LGCP's, \cite{Molleretal(98)} implement a Metropolis adjusted Langevin algorithm (MALA, \cite{Besag(94)}, \cite{RobertsTweedie(96b)} and \cite{RobertsRosenthal(98)}).
This algorithm achieves a higher asymptotic acceptance rate than a random walk Metropolis-Hastings algorithm (random walk MH, \cite{RobertCasella(04)}) by employing the transition density induced by the Langevin diffusion to the target distribution.
\cite{Diggleetal(13)} survey models of space-time and multivariate LGCP (mLGCP) and implement manifold MALA (MMALA, \cite{GirolamiCalderhead(11)}), which exploits the Riemann geometry of parameter spaces of models and achieve automatic adaptive tuning to the local structure through a preconditioning matrix (\cite{RobertsStramer(03)}).
More recently, \cite{LeiningerGelfand(16)} implement elliptical slice sampling proposed by \cite{MurrayAdamsMacKay(10)} for sampling high dimensional latent Gaussian variables of spatial LGCP.
This algorithm does not require the fine tuning and further computation of the target density information.
\cite{Chakrabortyetal(11)} discuss and introduce various ad hoc approaches used by ecologists in the context of species locations collected as so-called presence-only datasets.
They fit an LGCP with Gaussian predictive process approximation proposed by \cite{Banerjeeetal(08)}.
\cite{Pacietal(16)} construct space-time LGCP for the residential property sales data by implementing nearest neighbor GP recently proposed by \cite{Dattaetal(16a)}

Integrated nested Laplace approximation (INLA, \cite{Rueetal(09)}) has been proposed as an alternative Bayesian inference strategy.
It is a highly efficient approximate Bayesian inference scheme for structured latent GP models which approximate the marginal posterior distribution of the model hyperparameters by using Laplace approximation (\cite{TierneyKadane(86)}).
Approximating a Gaussian process or Gaussian random field (GRF) by a Gaussian Markov random field (GMRF, see \cite{RueHeld(05)}) results in a precision matrix rather than a covariance matrix.
No matrix inversion is required; computationally efficient evaluation of the posterior marginal distribution of hyperparameters and latent Gaussian variables is achieved.
\cite{Lindgrenetal(11)} show the connection of GMRF to GRF through a stochastic partial differential equation (SPDE), supporting the GMRF approximation.
Hence, they can estimate the hyperparameters of the GRF through this connection while utilizing the computationally efficient structure of the GMRF.
\cite{Illianetal(12)} implement INLA for inference with the LGCP, demonstrating its applicability with several ecological datasets.
More recently, \cite{Simpsonetal(16b)} implement INLA, employing the results in \cite{Lindgrenetal(11)}, for inference with the LGCP and discuss its convergence properties.
The INLA approach has been successfully implemented for the latent GP models where the dimension of the hyperparameters of the GP is low (basically 2 to 5, but not exceeding 20, see \cite{Rueetal(16)}).

Recently, Bayesian inference for the marginal posterior within the MCMC perspective using a so-called pseudo-marginal MCMC (PM-MCMC) approach has been proposed \citep{AndrieuRoberts(09)}.
The approach introduces an unbiased estimator of the marginal likelihood integrated over latent variables into the MH acceptance ratio instead of the likelihood itself.
The attractive and surprising property of the method is that convergence to the exact marginal posterior distribution is guaranteed when we use this unbiased estimator.
The efficiency of this algorithm is dependent on the variance of the unbiased estimator.
Hence, the primary task for this algorithm is to construct an unbiased estimator keeping its variance as small as possible.
A straightforward construction of this estimator is through importance sampling.
However, direct implementation of PM-MCMC for LGCP has a major computational problem, again, implementation of high dimensional importance sampling is required to construct an unbiased estimate.
Although the accuracy of the inference is dependent on the resolution of the approximation, the large grid approximation increases the variance of the unbiased estimator.

In this paper, we propose a general approximate Bayesian inference scheme for the LGCP.
This computational scheme is comprised of two steps.
At the first step, we estimate the approximate marginal posterior of the parameters by PM-MCMC.
To avoid the high dimensionality of the latent Gaussian variables, we take the grid approximation over the study region and convert the likelihood into the multivariate product Poisson log normal (mPLN, \cite{AitchisonHo(89)}).
At the second step, we calculate the marginal posterior of latent Gaussian variables.
Hence, our computational scheme is similar in spirit to INLA but without concern regarding the dimension of the hyperparameters.


The format of the paper is as follows.  Section 2 reviews some Bayesian inference approaches for LGCPs. Section 3 provides the development of our marginal posterior approximation for a LGCP along with the introduction of the pseudo marginal approach and kernel mixture marginalization. In Section 4, we investigate the proposed algorithms with some simulation investigation while in Section 5 we analyze tree species datasets.
Finally, Section 6 offers discussion regarding our approach.

\section{Reviewing Bayesian inference for log Gaussian Cox Processes}
\subsection{MCMC Based Inference}
Let $\mathcal{S}=\{\bm{s}_{1},\ldots, \bm{s}_{n}\}$ be the observed point pattern in $\mathcal{D} \subset \mathbb{R}^{d}$ is the study region. Then, given the intensity surface $\lambda(\cdot)$, the likelihood of LGCP is defined as
\begin{align}
\mathcal{L}(\mathcal{S}|\bm{\theta}, \bm{z})&=\exp\biggl(|D|-\int_{\mathcal{D}}\lambda(\bm{u}|\bm{\theta},z(\bm{u}))d\bm{u}\biggl)\prod_{\bm{s}\in \mathcal{S}}\lambda(\bm{s}|\bm{\theta},z(\bm{s})) \label{eq:like} \\
\log \lambda(\bm{s}|\bm{\theta},z(\bm{s}))&=\bm{X}(\bm{s})\bm{\beta}+z(\bm{s}), \quad \bm{z}(\mathcal{S})\sim \mathcal{N}(\bm{0}, \mathbf{C}_{\bm{\zeta}}(\mathcal{S}, \mathcal{S})), \quad \bm{s}\in \mathcal{D}
\end{align}
Here $\bm{X}(\bm{s})$ is a covariate vector with $\bm{\beta}$ an associated coefficient vector.  $\bm{z}(\mathcal{S})=\{z(\bm{s}_{1}), \ldots, z(\bm{s}_{n})\}$ denotes the $n$-variate zero mean Gaussian distributed variables with covariance matrix $\mathbf{C}_{\bm{\zeta}}(\mathcal{S},\mathcal{S})=(C_{\bm{\zeta}}(\bm{s}_{i}, \bm{s}_{j}))_{i,j=1,\ldots,n}$, with $\bm{\zeta}$ the vector for hyperparameters for the covariance function.
We collect all of the parameters as $\bm{\theta}$, i.e., $\bm{\theta}=(\bm{\beta},\bm{\zeta})$.

The stochastic integral inside the exponential is analytically intractable so approximation is required.
Finite grid approximation, using $K$ points, takes the form $\sum_{k=1}^{K}\lambda(\bm{u}_{k}|\bm{\theta},z(\bm{u}_{k}))\Delta_{k}$ where $\Delta_{k}$ is the area of $k$-th grid (\cite{Molleretal(98)} and \cite{Illianetal(12)}).
For this approximation, $K$ variables from the GP have to be sampled, so we need to calculate the inverse of a $K \times K$ covariance matrix within each MCMC iteration.

For sampling of the GP outputs and hyperparameters, one option is elliptical slice sampling (\cite{MurrayAdamsMacKay(10)} and \cite{MurrayAdams(10)}) as discussed in {\cite{LeiningerGelfand(16)}}.
Let $\bm{z}=\mathbf{L}_{\bm{\zeta}}\bm{\nu}$ where $\mathbf{C}_{\bm{\zeta}}=\mathbf{L}_{\bm{\zeta}}\mathbf{L}_{\bm{\zeta}}^{'}$ and $\bm{\nu}\sim \mathcal{N}(\bm{0}, \mathbf{I})$.
We draw proposal $\bm{\nu}^{*}$ for $\bm{\nu}$ along the elliptical curve, i.e., $\bm{\nu}^{*}=\bm{\nu} \cos(\omega)+\bm{\eta} \sin(\omega)$ where $\bm{\eta}\sim \mathcal{N}(\bm{0}, \mathbf{I})$ and $\omega \sim \mathcal{U}[0,2\pi)$ and $\bm{z}=\mathbf{L}_{\bm{\zeta}}\bm{\nu}^{*}$.
When $\log \mathcal{L}(\mathcal{S}|\bm{\theta}, \bm{z}^{*})>\log \mathcal{L}(\mathcal{S}|\bm{\theta}, \bm{z})+\log v$ where $v \sim \mathcal{U}[0,1]$, the proposal is accepted. Otherwise, the proposal region along the elliptical curve is shrunk and a new value is proposed.
This approach is easy to implement without tuning, $\bm{\nu}^{*}$ moves toward $\bm{\nu}$ automatically after rejection.
In this algorithm, sampling the hyperparameters does not involve direct evaluation of the prior distribution of the GP ($\mathcal{N}(\bm{z}(\mathcal{S})|\bm{0},\mathbf{C}_{\bm{\zeta}}(\mathcal{S},\mathcal{S}))$) due to the transformation of variables (\cite{Molleretal(98)} and \cite{MurrayAdams(10)}).
Hence, the calculation of the inverse of covariance matrix is not required but Cholesky decomposition is still necessary.

The Metropolis adjusted Langevin algorithm (MALA) was originally suggested by \cite{Besag(94)} and further investigated by \cite{RobertsTweedie(96b)} and \cite{RobertsRosenthal(98)}.
This approach is a Metropolis-Hastings (MH) algorithm with the transition density driven by Langevin diffusion, $\bm{z}^{*}=\bm{z}^{(i-1)}+\sigma_{n}\bm{\nu}^{(i-1)}+\frac{\sigma_{n}^2}{2}\nabla \log\{\pi(\bm{z}^{(i-1)}|\mathcal{S}) \}$ where the random variables $\bm{\nu}^{(i-1)}$ are distributed as independent standard normal and $\sigma_{n}^2$ is the step variance.
Since the Langevin algorithm utilizes local gradient information of the target density, a higher optimal acceptance rate is accomplished than with random walk MH (\cite{RobertsRosenthal(98)}).
Although the drift term in the proposal for the MALA introduces the direction for the proposal based on the gradient information, the isotropic diffusion will be inefficient for strongly correlated variables (\cite{GirolamiCalderhead(11)}).
For this problem, \cite{RobertsStramer(03)} introduce a {\it preconditioning matrix} $\mathbf{M}$ such that $\bm{z}^{*}=\bm{z}^{(i-1)}+\sigma_{n}\mathbf{M}^{-1/2}\bm{\nu}^{(i-1)}+\frac{\sigma_{n}^2}{2}\mathbf{M}^{-1}\nabla \log\{\pi(\bm{z}^{(i-1)}|\mathcal{S}) \}$.
However, it is unclear how to specify  $\mathbf{M}$.
\cite{GirolamiCalderhead(11)} propose manifold MALA (MMALA), which incorporates the Riemann geometry of the parameter spaces of the models in order to achieve automatic adaptive tuning to the local structure through a preconditioning matrix and investigate algorithms for various applications including LGCP.
For a LGCP, $\mathbf{M}$ is a $K\times K$ full matrix, the computational cost of inversion of this preconditioning matrix is $\mathcal{O}(K^3)$.
So, although their approach achieves sampling efficiency, iterative calculation of a preconditioning matrix for a LGCP is computationally costly.
In summary, these foregoing MCMC based inference approaches for LGCPs
introduce computational cost $\mathcal{O}(K^3)$ and $\mathcal{O}(K^2)$ memory for storage.

\subsection{Integrated Nested Laplace Approximation}

Integrated nested Laplace approximation (INLA) proposed by \cite{Rueetal(09)} offers highly efficient approximate Bayesian inference for structured latent Gaussian models.
A very recent review is \cite{Rueetal(16)}. We briefly review INLA here to facilitate comparison with the approximation approach we propose in Section 3.
Let $\bm{y}=\{y_{1},\ldots, y_{n}\}$ be an observed dataset, modeled as $\pi(\bm{y}|\bm{\beta}, \bm{z})$ with $\bm{z}$ modeled as $\pi(\bm{z}|\bm{\zeta})$. The approach approximates the marginal posterior of latent Gaussian variables.
In particular, INLA implements Laplace approximation (\cite{TierneyKadane(86)} and \cite{BarndorffNielsenCox(89)}) for estimating these posteriors.  
With the approximate posterior for $\bm{\zeta}$, $\tilde{\pi}(\bm{\zeta}|\bm{y})$,
the approximate marginal posterior of $z_{i}$ for $i=1,\ldots, n$ is evaluated as
\begin{align}
\tilde{\pi}(z_{i}|\bm{y})=\sum_{j} \tilde{\pi}(z_{i}|\bm{\zeta}_{j},\bm{y})\tilde{\pi}(\bm{\zeta}_{j}|\bm{y})\Delta_{\bm{\zeta}_{j}}.
\end{align}
where $\Delta_{\bm{\zeta}_{j}}$ is the area of the grid cell for $\bm{\zeta}_{j}$.
The sum is over the gridded values of $\bm{\zeta}_{j}$ with area $\Delta_{\bm{\zeta}_{j}}$. 

Calculation of $\tilde{\pi}(\bm{\zeta}|\bm{y})$ arises as
\begin{align}
\tilde{\pi}(\bm{\zeta}|\bm{y})\propto \frac{\pi(\bm{z},\bm{y},\bm{\zeta})}{\tilde{\pi}(\bm{z}|\bm{\zeta},\bm{y})}\biggl|_{\bm{z}=\bm{z}^{*}(\bm{\zeta})}
\end{align}
where $\tilde{\pi}(\bm{z}|\bm{\zeta},\bm{y})$ is the Gaussian approximation to the full conditional of $\bm{z}$, and $\bm{z}^{*}(\bm{\zeta})$ is the mode of the full conditional for $\bm{z}$ given $\bm{\zeta}$ (\cite{Rueetal(09)} and \cite{Martinsetal(13)}).
In the context of a Gaussian Markov random field (GMRF, \cite{RueHeld(05)}), Laplace approximation for precision matrices of GP yields fast computation because the zeros in the precision matrix imply conditional independence of the variables. As a result, the computational cost for the spatial GMRF case is $\mathcal{O}(K^{3/2})$ computational time with $\mathcal{O}(K\log (K))$ memory for storage (see \cite{RueHeld(05)}).

\cite{Illianetal(12)} implement the INLA approach for a LGCP and \cite{Simpsonetal(16b)} investigate its convergence properties using the theoretical results regarding the connection between the GMRF and GRF in \cite{Lindgrenetal(11)}.
Although INLA based inference for a LGCP is highly efficient, there are potential problems.
First, the INLA approach has to evaluate $\tilde{\pi}(\bm{\zeta}_{j}|\bm{y})$ for each gridded $\bm{\zeta}_{j}$ and then integrate over $\pi(\bm{\zeta}|\bm{y})$ to calculate $\tilde{\pi}(z_{i}|\bm{y})$.
For example, if we take three integration points in each dimension, the cost would be $3^{p}$ to cover all combinations in $p$ dimensions, which is 729 for $p=6$ and 59049 for $p=10$.
Hence, INLA based inference is really practical only for low dimensional $\bm{\theta}$ along with coarse grids (see, \cite{Rueetal(09)}, \cite{Illianetal(12)}, and \cite{Simpsonetal(16b)}).
Point pattern models requiring more complex covariance specification, e.g., nonstationarity, nonseparability in space and time and multivariate spatial processes may exceed the capability of INLA.
For this problem, \cite{Rueetal(09)} and \cite{Martinsetal(13)} consider central composite design (CCD) integration, which designs the evaluation points by using the mode $\bm{\theta}^{*}$ and the negative Hessian matrix as a guide under the Gaussian assumption.
The method is a default setting for taking evaluation points in the case where the number of hyperparameters is larger than two in the \texttt{R-INLA} package (\cite{LindgrenRue(15)}).
However, Gaussian assumptions with negative Hessian matrix as a guide might not be accurate enough when the marginal posterior distribution of some components of $\bm{\zeta}$ is strongly skewed, e.g., a variance parameter in a covariance matrix.


Furthermore, the INLA approach evaluates the posterior component-wise, i.e., $\pi(z_{i}|\bm{\zeta}, \bm{y})$.
This specification does not enable us to evaluate joint posterior probability or posterior distribution of nonlinear functionals of $\bm{z}$, e.g., the maximum of components of $\bm{z}$ (\cite{Diggle(09)}) and gradients of  the intensity surface (\cite{Liangetal(09)}).
Such nonlinear functionals of $\bm{z}$ have practical interest, e.g., hotspot detection in crime event data (\cite{Bowersetal(04)}) or cancer risk factors (\cite{Liangetal(09)}).
\section{Pseudo-marginal  for exact MCMC for the approximate marginal posterior of the parameters}
Here, we lay out a computational scheme based on pseudo-marginal MCMC originally introduced by \cite{Beaumont(03)} and further investigated by \cite{AndrieuRoberts(09)}.
Since our computational scheme is specific for a LGCP, we assume data $\mathcal{S}$, i.e., an observed point pattern.
Our approach has a flavor similar to that of INLA, we seek efficient and accurate Bayesian inference under an approximate marginal posterior distribution for $\bm{\theta}=(\bm{\beta},\bm{\zeta})$, i.e., $\tilde{\pi}(\bm{\theta}|\mathcal{S})$.
This approximate marginal posterior distribution is constructed in a different way from INLA, tuned for the LGCP setting, and is estimated in the MCMC framework.
After obtaining this approximate marginal posterior, we can obtain the joint marginal posterior of the latent Gaussian variables as
\begin{align}
\tilde{\pi}(\bm{z}|\mathcal{S})=\sum_{i=i_{0}+1}^{I+i_{0}}\pi(\bm{z}|\bm{\theta}^{(i)},\mathcal{S})\tilde{\pi}(\bm{\theta}^{(i)}|\mathcal{S})
\end{align}
where $I$ is the number of approximate marginal posterior samples and $i_{0}$ is the end point of burn-in period.

Given the retained $\{\bm{\theta}^{(i)}\}_{i=i_{0}+1}^{I+i_{0}}$, we could estimate $\pi(\bm{z}|\bm{\theta}^{(i)},\mathcal{S})$ at each $\bm{\theta}^{(i)}$.
Since we can sample $\pi(\bm{z}|\bm{\theta}^{(i)},\mathcal{S})$ at fixed $\bm{\theta}^{(i)}$, calculation of inverse or Cholesky decomposition of covariance matrices can be implemented independently with respect to different $\bm{\theta}^{(i)}$.
As long as multiple cores/nodes and computers are available, we can separate the estimation tasks into each core or computer without any further algorithms. So the heavy iteration through MCMC, which is not parallelizable, is not required.
In practice, we might take thinning over $I$ approximate marginal posterior samples of $\bm{\theta}$, e.g., so that $\bm{\theta}$ consists of approximately independent samples from the approximate marginal posterior distribution.
Furthermore, since $\bm{\theta}^{(i)}$ is fixed, $\bm{z}$ converges more rapidly to $\pi(\bm{z}|\bm{\theta}^{(i)},\mathcal{S})$ (\cite{Filipponeetal(13)}).
The sampling inefficiency caused by the high correlation between $\bm{z}$ and $\bm{\theta}$ is mitigated.
The posterior samples of $\bm{z}$ are obtained by elliptical slice sampling for LGCP (\cite{LeiningerGelfand(16)}) or MALA (\cite{Molleretal(98)} and \cite{Diggleetal(13)}).
This step can be implemented independently with respect to different $\bm{\theta}^{(i)}$, since we only need to sample $\bm{z}|\mathcal{S}, \bm{\theta}^{(i)}$ at fixed $\bm{\theta}^{(i)}$. Hence, we turn to the question of how to accurately estimate $\pi(\bm{\theta}|\mathcal{S})$ for an LGCP.

For working with latent variables, the pseudo-marginal approach was introduced by \cite{Beaumont(03)} and its convergence properties were investigated by \cite{AndrieuRoberts(09)}.
The key point is to construct an unbiased estimate of the likelihood, in our case $\hat{\pi}(\mathcal{S}|\bm{\theta})$, i.e., the likelihood by integrating out latent variables. Then, this estimated likelihood is put into the Metropolis Hastings (MH) acceptance ratio, i.e.,
\begin{align}
\alpha=\biggl\{1,\frac{\pi(\bm{\theta}^{*})\hat{\pi}(\mathcal{S}|\bm{\theta}^{*})q(\bm{\theta}|\bm{\theta}^{*})}{\pi(\bm{\theta})\hat{\pi}(\mathcal{S}|\bm{\theta})q(\bm{\theta}^{*}|\bm{\theta})} \biggl\}
\end{align}
If a uniform random variable $u$ is drawn and $u<\alpha$, we retain $\bm{\theta}^{*}$ and $\hat{\pi}(\mathcal{S}|\bm{\theta}^{*})$.
Surprisingly, the convergence to $\pi(\bm{\theta}|\mathcal{S})$ is guaranteed as long as $\hat{\pi}(\mathcal{S}|\bm{\theta})$ is an unbiased estimate of $\pi(\mathcal{S}|\bm{\theta})$.
\cite{AndrieuRoberts(09)} demonstrate the uniform ergodicity of the algorithm.
The efficiency is dependent on the variance of $\hat{\pi}(\mathcal{S}|\bm{\theta})$.
When $\hat{\pi}(\mathcal{S}|\bm{\theta})$ is a noisy estimate, the accepted samples will be highly correlated.
So, the main task for the pseudo-marginal approach is to construct an unbiased estimate $\hat{\pi}(\mathcal{S}|\bm{\theta})$ with small variance over $\bm{\theta}$.

The straightforward approach is importance sampling (e.g., \cite{RobertCasella(04)}) which enables construction of an unbiased estimate with smaller variance than the direct Monte Carlo estimate.
\cite{FilipponeGirolami(14)} implement the pseudo-marginal approach for estimating hyperparameters of GP.
They utilize Laplace approximation (\cite{TierneyKadane(86)} and \cite{BarndorffNielsenCox(89)}) and expectation propagation (\cite{Minka(01)}) as the importance density to construct an unbiased estimate.
Although their approach itself doesn't assume a specific form for the likelihood, the number of samples is relatively small (less than 3,000) due to the computational costs of importance sampling.
Furthermore, pseudo-marginal approach achieves its attractive property under MCMC; parallelization of the algorithm is not straightforward without approximation.
For the LGCP, we need to introduce a grid approximation over the study region.
For an accurate implementation, sufficiently fine grids are required with regard to approximating the infinite dimensional stochastic integral.
However, importance sampling for our high dimensional case is not promising.

Let $\bm{B}=(B_{1},\ldots, B_{M})$ be $M$ disjoint partitions over $\mathcal{D}$, i.e., $\bigcup_{m=1}^{M}B_{m}=\mathcal{D}$, with $\bm{T}(\mathcal{S})=(T_{1}(\mathcal{S}),\ldots,T_{M}(\mathcal{S}))$ and $\bm{\delta}=(\delta_{1},\ldots, \delta_{M})$ the counts and intensities, respectively on each subregion.
That is, direct pseudo-marginal MCMC assumes a homogeneous Poisson on each grid as in standard MCMC and in INLA approaches.
More specifically, we take $\bm{T}(\mathcal{S})$ as a set of \emph{summary} statistics for $\mathcal{S}$ and construct an approximate distribution for $\bm{T}(\mathcal{S})$ given $\bm{\theta}$.  Below, we take $\bm{T}(\mathcal{S})$ as the counts associated with the $B_{m}$s and utilize the multivariate Poisson log normal (mPLN) kernel function (\cite{AitchisonHo(89)}) to approximate this distribution. Then, with a prior on $\bm{\theta}$, we take $\tilde{\pi}(\bm{\theta}|\mathcal{S}) =\pi(\bm{\theta}|\bm{T}(\mathcal{S}))$.

The direct implementation of the pseudo-marginal approach requires high dimensional grid approximation over $\mathcal{D}$, i.e., integrating out $M$ latent Gaussian variables where $M$ is large. The estimator is given by
\begin{align}
\hat{\pi}(\bm{T}(\mathcal{S})|\bm{\theta})=\frac{1}{N_{imp}}\sum_{j=1}^{N_{imp}}\frac{\pi(\bm{T}(\mathcal{S})|\bm{\delta}_{j})\pi(\bm{\delta}_{j}|\bm{\theta})}{g(\bm{\delta}_{j}|\bm{T}(\mathcal{S}),\bm{\theta})}, \quad \bm{\delta}_{j}\sim g(\bm{\delta}_{j}|\bm{T}(\mathcal{S}), \bm{\theta})
\end{align}
where $N_{imp}$ is the number of samples from the importance density $g(\bm{\delta}|\bm{T}(\mathcal{S}), \bm{\theta})$. When $M$ is large (e.g., $M=K$), obtaining an estimate with small variance is computationally demanding because a very large $N_{imp}$ is needed.
The straightforward implementation requires large $M$ due to the "local" homogeneity approximation for the intensity.

In our approach, we avoid high dimensional integration by utilizing the first and second moment equations induced by general moment formula for Cox processes. We can calculate the exact first and second order moments from the general moments formula for Cox processes (see below).
So, although we take grid approximation as in the direct implementation, for us, the first and second order moments of the distribution for $\bm{T}$ are induced by the moments of the LGCP. These moment choices eliminate biases caused by the grid approximation (homogeneous Poisson on each grid) in INLA and MCMC based algorithms with insufficient grids.
\subsection{Kernel mixture marginalization}

We consider kernel mixture marginalization for the density of $\bm{T}$,  i.e.,
\begin{align}
\pi(\bm{T}(\mathcal{S})|\bm{\theta})=\int \prod_{m=1}^{M}\mathcal{P}(T_{m}(\mathcal{S})|\delta_{m})\pi(\bm{\delta}|\bm{\theta})d\bm{\delta}
\label{eq:margT}
\end{align}
where $\pi(\bm{\delta}|\bm{\theta})$ is the prior distribution of the intensity vector and $\mathcal{P}(T_{m}(\mathcal{S})|\delta_{m})$ is Poisson distribution with intensity $\delta_{m}$.
This mixture representation incorporates the correlation structure for the counts into the intensity distribution.
That is, conditionally, given the intensity, the counts for different grid cells are independent with a product Poisson joint distribution but marginally, they are dependent.
The direct numerical integration for equation (\ref{eq:margT}) is difficult when $M$ is large.
In our approach, we introduce the exact first and second order moments induced from Cox processes to keep $M$ is small ($<K$).

From the moment formula for general Cox processes (e.g., \cite{MollerWaagepetersen(04)}), we can obtain the first and second moments of the marginal counts summary vector given $\bm{\theta}$, i.e., $\alpha_{\bm{\theta},m}=E[T_{m}(\mathcal{S})|\bm{\theta}]$ and $\beta_{\bm{\theta},mm'}=Cov[T_{m}(\mathcal{S}), T_{m'}(\mathcal{S})|\bm{\theta}]$ for $m,m'=1,\ldots, M$,
\begin{align}
\alpha_{\bm{\theta},m}&=\int_{B_{m}}\lambda_{\bm{\theta}}(\bm{u})d\bm{u}, \quad \lambda_{\bm{\theta}}(\bm{s})=E_{\bm{z}}[\lambda(\bm{s}|\bm{\theta},z(\bm{s}))] \quad \bm{s} \in \mathcal{D} \\
\beta_{\bm{\theta}, mm'}&=\int_{B_{m}\cap B_{m'}}\lambda_{\bm{\theta}}(\bm{u})d\bm{u}+\int_{B_{m}}\int_{B_{m'}}\lambda_{\bm{\theta}}(\bm{u})\lambda_{\bm{\theta}}(\bm{v})\{g_{\bm{\theta}}(\bm{u},\bm{v})-1\}d\bm{u}d\bm{v},
\end{align}
where $g_{\bm{\theta}}(\bm{u}, \bm{v})$ is the pair correlation function of the latent process.
Since $B_{m}$ and $B_{m'}$ are disjoint, $\int_{B_{m}\cap B_{m'}}\lambda_{\bm{\theta}}(\bm{u})d\bm{u}=\int_{B_{m}}\lambda_{\bm{\theta}}(\bm{u})d\bm{u}$ for $m=m'$ and 0 otherwise. $\lambda_{\bm{\theta}}(\bm{s})=E_{\bm{z}}[\lambda(\bm{s}|\bm{\theta}, z(\bm{s}))]$ is the expected intensity with respect to $z$. For the LGCP, the expected intensity $\lambda_{\bm{\theta}}(\bm{s})$ and pair correlation function $g_{\bm{\theta}}(\bm{u}, \bm{v})$ are obtained analytically as $\lambda_{\bm{\theta}}(\bm{s})=\exp(\bm{X}(\bm{s})\bm{\beta}+\sigma^2(\bm{s})/2)$ and $g_{\bm{\theta}}(\bm{u}, \bm{v})=\exp(C_{\bm{\zeta}}(\bm{u}, \bm{v}))$, where $\sigma^2(\bm{s})=C_{\bm{\zeta}}(\bm{s}, \bm{s})$, in practice we assume $\sigma^2(\bm{s})=\sigma^2$ (\cite{Molleretal(98)}).

Although the marginal density $\pi(\bm{T}(\mathcal{S})|\bm{\theta})$ in (\ref{eq:margT}) is not analytically available, the first and second order moments can be calculated from above formulas for any moment assumptions for the latent process. In practice, these moments might not be available explicitly but can be accurately calculated by grid approximation, i.e.,
\begin{align}
\hat{\alpha}_{\bm{\theta},m}&=\sum_{b=1}^{N_{B_{m}}}\lambda_{\bm{\theta}}(\bm{u}_{b})\Delta_{B_{m}} \label{eq:alphahat}, \quad  \bm{u}_{b} \in B_{m}, \quad \bm{v}_{b'} \in B_{m'} \\
\hat{\beta}_{\bm{\theta},mm'}&=\sum_{b=1}^{N_{B_{m}\cap B_{m'}}}\lambda_{\bm{\theta}}(\bm{u}_{b})\Delta_{B_{m}\cap B_{m'}}+\sum_{b=1}^{N_{B_{m}}}\sum_{b^{'}=1}^{N_{B_{m'}}}\lambda_{\bm{\theta}}(\bm{u}_{b})\lambda_{\bm{\theta}}(\bm{v}_{b^{'}})\{g_{\bm{\theta}}(\bm{u}_{b}, \bm{v}_{b^{'}})-1\}\Delta_{B_{m}}\Delta_{B_{m'}} \label{eq:betahat}
\end{align}
where $\{\bm{u}_{b} \}_{b=1}^{N_{B_{m}}}$, $N_{B_{m}}$ and $\Delta_{B_{m}}$ are representative points, the number of grids and the area of a grid within $B_{m}$.
For simplicity, we assume $N_{B_{m}}=N_{B}$ and $\Delta_{B_{m}}=\Delta_{B}$ for all $m$.
Importantly, $\bm{B}$ is an $M$-dimensional  vector of disjoint grid cells over the study region used to construct $\bm{T}$.  However, the above moment calculation is implemented by taking $\it{further}$ grid cells within each $B_{m}$.
That is, we take an $M$-dimensional count summary statistics vector over the $M$-dimensional disjoint subregions at first stage. Then, we calculate exact first and second order moments of $\bm{T}|\bm{\theta}$ on $\bm{B}$ through the above moments formula directly.

Now, we specialize (\ref{eq:margT}) to obtain the product Poisson log normal kernel version by introducing a log normal intensity function arising from the LGCP,
\begin{align}
\pi(\bm{T}(\mathcal{S})|\bm{\theta})=\int \prod_{m=1}^{M}\mathcal{P}(T_{m}(\mathcal{S})|\delta_{m})\mathcal{N}(\log \bm{\delta}|\bm{\mu}_{\bm{\theta}}, \mathbf{\Sigma}_{\bm{\theta}})d\bm{\delta}
\end{align}
where $\bm{\mu}_{\bm{\theta}}=(\mu_{\bm{\theta},1},\ldots,\mu_{\bm{\theta},M})$ and $\mathbf{\Sigma}_{\bm{\theta}}=(\sigma_{\bm{\theta},mm'})_{m,m'=1,\ldots, M}$ are the mean vector and covariance matrix of $\log (\bm{\delta})$. In this approach, the latent intensity parameter $\bm{\delta}$ is introduced and the marginal correlation structure is incorporated into the log normal distribution of $\bm{\delta}$. The resulting marginal mean and variance structure of $\bm{T}(\mathcal{S})|\bm{\theta}$ (see, \cite{AitchisonHo(89)}) yields the system of equations
\begin{align}
\alpha_{\bm{\theta},m}&=\exp\biggl(\mu_{\bm{\theta}, m}+\frac{\sigma_{\bm{\theta},mm}}{2}\biggl), \quad m=1,\ldots, M \\
\beta_{\bm{\theta},mm'}&=\bm{1}(m=m^{'})\alpha_{\bm{\theta}, m}+\alpha_{\bm{\theta}, m}\alpha_{\bm{\theta}, m'}\{\exp(\sigma_{\bm{\theta},mm'})-1\}, \quad m, m'=1,\ldots, M
\end{align}
We can invert this system of equations, that is, $\bm{\mu}_{\bm{\theta}}$ and $\mathbf{\Sigma}_{\bm{\theta}}$ are induced from $\bm{\alpha}_{\bm{\theta}}$ and $\bm{\beta}_{\bm{\theta}}$ as
\begin{align}
\mu_{\bm{\theta},m}&=\log (\alpha_{\bm{\theta}, m})-\frac{\sigma_{\bm{\theta},mm}}{2} \\
\sigma_{\bm{\theta},mm}&=\log \biggl(1+\frac{\beta_{\bm{\theta}, mm}}{\alpha_{\bm{\theta}, m}^2}-\frac{1}{\alpha_{\bm{\theta}, m}}\biggl), \quad
\sigma_{\bm{\theta},mm'}=\log \biggl(1+\frac{\beta_{\bm{\theta}, mm'}}{\alpha_{\bm{\theta}, m}\alpha_{\bm{\theta}, m'}} \biggl)
\end{align}
Since $\beta_{\bm{\theta},mm'}$ for $m' \neq m$ can be positive and negative, both positive and negative correlation among counts can be incorporated.
However, this specification expresses only overdispersion because the marginal variance $\beta_{\bm{\theta},mm}$ has to be larger than $\alpha_{\bm{\theta},m}$ to satisfy $\sigma_{\bm{\theta},mm}>0$.
Evidently, the total number of parameters in $\mu_{\bm{\theta}}$ and $\mathbf{\Sigma}_{\bm{\theta}}$ is $\frac{1}{2}M(M+3)$,  the same number as that in $\bm{\alpha}_{\bm{\theta}}$ and $\bm{\beta}_{\bm{\theta}}$.
Hence, the matching between $(\bm{\alpha}_{\bm{\theta}}, \bm{\beta}_{\bm{\theta}})$ and $(\bm{\mu}_{\bm{\theta}}, \mathbf{\Sigma}_{\bm{\theta}})$ is one to one.
Actually, $\sigma_{\bm{\theta},mm}$ and $\sigma_{\bm{\theta},mm^{'}}$ are deterministic functions of ($\alpha_{\bm{\theta}, m}$, $\alpha_{\bm{\theta}, m^{'}}$, $\beta_{\bm{\theta}, mm}$, $\beta_{\bm{\theta}, mm^{'}}$), which are obtained from equation (\ref{eq:alphahat}) and (\ref{eq:betahat}).

Since the expressions for $\bm{\alpha}_{\bm{\theta}}$ and $\bm{\beta}_{\bm{\theta}}$ are exact, the induced moments $\bm{\mu}_{\bm{\theta}}$ and $\mathbf{\Sigma}_{\bm{\theta}}$ are also the exact mean and covariance for $\log \bm{\delta}$ under this model.
To summarize, at first, we need to calculate $\bm{\alpha}_{\bm{\theta}}$ and $\bm{\beta}_{\bm{\theta}}$ which are dependent on $\bm{\theta}$.  Then, we transform these values into $\bm{\mu}_{\bm{\theta}}$ and $\mathbf{\Sigma}_{\bm{\theta}}$.
\subsection{Construction of the unbiased estimator}
Based on the above discussion, we construct an unbiased estimator in our context.
The estimator is given by
\begin{align}
\hat{\pi}(\bm{T}(\mathcal{S})|\bm{\theta})=\frac{1}{N_{imp}}\sum_{j=1}^{N_{imp}}\frac{\prod_{m=1}^{M}\mathcal{P}(T_{m}(\mathcal{S})|\delta_{m,j})\mathcal{N}(\log \bm{\delta}_{j}|\bm{\mu}_{\bm{\theta}}, \mathbf{\Sigma}_{\bm{\theta}})}{g(\bm{\delta}_{j}|\bm{T}(\mathcal{S}),\bm{\theta})}, \quad \bm{\delta}_{j}\sim g(\bm{\delta}_{j}|\bm{T}(\mathcal{S}), \bm{\theta})
\label{eq:UE}
\end{align}
where $N_{imp}$ is the number of samples from the importance density $g(\bm{\delta}|\bm{T}(\mathcal{S}), \bm{\theta})$.
Through the theory of importance sampling (\cite{RobertCasella(04)}), an unbiased estimator created using suitable importance sampling can have smaller variance than a direct Monte Carlo estimator.

As for the specification of the importance density $g(\bm{\delta}|\bm{T}(\mathcal{S}), \bm{\theta})$, we employ Laplace approximation for the posterior distribution of $\bm{\delta}$ as a default choice. In searching for a mode for the numerator, we implement a gradient descent algorithm as in \cite{Molleretal(98)}.
An alternative approach is expectation propagation (EP, \cite{Minka(01)}). \cite{FilipponeGirolami(14)} implement EP in addition to Laplace approximation, suggesting that EP might be more robust to the choice of $N_{imp}$. However, EP includes the iteration steps for updating the density, so computation might become to demanding.
Below, we adopt Laplace approximation.

Although the INLA approach requires fine grids over the study region, for some grid cells, the counts will be small.
Since Laplace approximation for the small counts will not be well approximated using a Gaussian distribution, a skewness corrected method is considered in INLA (\cite{Martinsetal(13)}).
With our approach we can potentially obtain relatively larger counts in each grid cell because $M$ can be smaller than in INLA. 

\subsection{The fitting algorithm}
We call the algorithm below the approximate marginal posterior (AMP) approach.  It is comprised of two steps - inference about $\bm{\theta}$ and then about $\bm{z}$.  Inference about $\bm{z}$ enables insight into the local behavior of the intensity surface of the LGCP, where it pushes up or pulls down the intensity to better explain the data.
So, the second step can be skipped if inference about the marginal posterior distribution for $\bm{z}$ is not of interest.
\\
\noindent{\bf Estimation of $\tilde{\pi}(\bm{\theta}|\mathcal{S})$}
\begin{itemize}
\item[1.] Let $i=1$, set initial value $\bm{\theta}^{(0)}$,
\item[2.] Generate $\bm{\theta}^{*}\sim q(\bm{\theta}|\bm{\theta}^{(i-1)})$, calculate moments
 $\bm{\alpha}_{\bm{\theta}^{*}}=E[\bm{T}(\mathcal{S})|\bm{\theta}^{*}]$ and $\bm{\beta}_{\bm{\theta}^{*}}=Cov[\bm{T}(\mathcal{S})|\bm{\theta}^{*}]$ and convert these moment vectors into $\bm{\mu}_{\bm{\theta}^{*}}$ and $\mathbf{\Sigma}_{\bm{\theta}^{*}}$.
\item[3.] Calculate the Laplace approximation $g(\bm{\delta}|\bm{T}(\mathcal{S}), \bm{\theta}^{*})$ for $\pi(\bm{T}(\mathcal{S})|\bm{\delta})\pi(\bm{\delta}|\bm{\theta}^{*})$.
\item[4.] Estimate $\hat{\pi}(\bm{T}(\mathcal{S})|\bm{\theta}^{*})$ as $\hat{\pi}(\bm{T}(\mathcal{S})|\bm{\theta}^{*})=\frac{1}{N_{imp}}\sum_{j=1}^{N_{imp}}\frac{\pi(\bm{T}(\mathcal{S})|\bm{\delta}_{j})\pi(\bm{\delta}_{j}|\bm{\theta}^{*})}{g(\bm{\delta}_{j}|\bm{T}(\mathcal{S}), \bm{\theta}^{*})}
$ where $g(\bm{\delta}_{j}|\bm{T}(\mathcal{S}), \bm{\theta}^{*})$ is the Laplace approximation of $\pi(\bm{T}(\mathcal{S})|\bm{\delta})\pi(\bm{\delta}|\bm{\theta}^{*})$ evaluated at $\bm{\delta}_{j}$ and $\bm{\delta}_{j}\sim g(\bm{\delta}|\bm{T}(\mathcal{S}), \bm{\theta}^{*})$ for $j=1,\ldots, N_{imp}$.
\item[5.] Evaluate the acceptance ratio $\alpha=\text{min}\biggl\{1, \frac{\pi(\bm{\theta}^{*})\hat{\pi}(\bm{T}(\mathcal{S})|\bm{\theta}^{*})q(\bm{\theta}^{(i-1)}|\bm{\theta}^{*})}{\pi(\bm{\theta}^{(i-1)})\hat{\pi}(\bm{T}(\mathcal{S})|\bm{\theta}^{(i-1)})q(\bm{\theta}^{*}|\bm{\theta}^{(i-1)})} \biggl\}
$ and retain $\bm{\theta}^{(i)}=\bm{\theta}^{*}$ and $\hat{\pi}(\bm{T}(\mathcal{S})|\bm{\theta}^{(i)})=\hat{\pi}(\bm{T}(\mathcal{S})|\bm{\theta}^{*})$ if $u<\alpha$ where $u\sim \mathcal{U}(0,1)$, otherwise $\bm{\theta}^{(i)}=\bm{\theta}^{(i-1)}$ and $\hat{\pi}(\bm{T}(\mathcal{S})|\bm{\theta}^{(i)})=\hat{\pi}(\bm{T}(\mathcal{S})|\bm{\theta}^{(i-1)})$. Back to step 2 and $i\to i+1$.
\end{itemize}

So, pseudo marginal MCMC enables us to estimate $\pi(\bm{\theta}|\bm{T}(\mathcal{S}))$.
Again, we see a difference between the AMP approach and other grid approximation based approaches.  For example, INLA and MALA assume (sample) a homogeneous Poisson variable on each grid cell, so they require high dimensional (sufficiently fine) gridding for accurate inference.
Instead, the AMP approach considers the intensity on broader grid cells whose distribution is induced from exact first and second order moments of Cox processes.  \\


\noindent{\bf Estimation of $\pi(\bm{z}|\bm{\theta}, \mathcal{S})$ and $\pi(\bm{z}|\mathcal{S})$ (optional)}
\begin{itemize}
\item[1.] Given $\bm{\theta}^{(i)}$ for $i=i_{0}+1,\ldots, I+i_{0}$, where $i_{0}$ is the end point of burn-in period and $I$ is the number of preserved approximate marginal posterior samples, estimate $\pi(\bm{z}|\bm{\theta}^{(i)}, \mathcal{S})$.
\item[2.] Calculate $\tilde{\pi}(\bm{z}|\mathcal{S})$ as $\tilde{\pi}(\bm{z}|\mathcal{S})=\sum_{i=i_{0}+1}^{I+i_{0}}\pi(\bm{z}|\bm{\theta}^{(i)}, \mathcal{S})\tilde{\pi}(\bm{\theta}^{(i)}|\mathcal{S})
$
\end{itemize}
There are four important points here.
First, since we retained $\{\bm{\theta}^{(i)}\}_{i=i_{0}+1}^{I+i_{0}}$ through the first step, we can calculate $\pi(\bm{z}|\bm{\theta}^{(i)}, \mathcal{S})$ separately for each $\bm{\theta}^{(i)}$.  We avoid the computational bottleneck of MCMC because
we need just a single Cholesky decomposition of the covariance matrix $\mathbf{C}_{\bm{\zeta}}$ for each $\bm{\theta}^{(i)}=(\bm{\beta}^{(i)}, \bm{\zeta}^{(i)})$.
These calculations can be implemented independently without passing of information.

Second, since iterative inversion or Cholesky decomposition is not required, we can implement them for relatively large covariance matrices without approximation.
Given fixed $\bm{\theta}^{(i)}$, convergence of $\pi(\bm{z}|\bm{\theta}^{(i)}, \mathcal{S})$ is fast say, using elliptical slice sampling (\cite{Filipponeetal(13)}).

Third, the AMP approach does not require grid approximation over $\bm{\theta}$. This can be a fatal bottleneck in the extension of the INLA approach to relatively larger dimensional hyperparameters $\bm{\zeta}$.
Although $\tilde{\pi}(\bm{\theta}|\mathcal{S})$ is still approximation of $\pi(\bm{\theta}|\mathcal{S})$ in the meaning of $\tilde{\pi}(\bm{\theta}|\mathcal{S})=\pi(\bm{\theta}|\bm{T}(\mathcal{S}))$, the marginal posterior distribution is well estimated as shown in simulation studies.
The AMP approach is potentially advantageous for larger dimensional $\bm{\zeta}$ (as a result, for larger dimensional $\bm{\theta}$) compared with the INLA approach.

Finally, since we estimate $\pi(\bm{z}|\bm{\theta},\mathcal{S})$ and $\pi(\bm{z}|\mathcal{S})$ via a sample based approach, the joint posterior distribution of $\bm{z}$ as well as the posterior distribution of nonlinear functionals of $\bm{z}$ are available.
\subsection{Computational Costs and Tuning Parameters}
There are three main computational costs for implementing the AMP algorithm.
First, the first and second order moments of $\bm{T}(\mathcal{S})|\bm{\theta}$, i.e., $\bm{\alpha}_{\bm{\theta}}$ and $\bm{\beta}_{\bm{\theta}}$, have to be calculated for each proposed $\bm{\theta}^{*}$.
Particularly, calculation of $\bm{\beta}_{\bm{\theta}}$ would be time consuming because the number of components is $M(M+1)/2$. So, the computational cost for these moment calculation is $\mathcal{O}(N_{B}M^2)$.
On the other hand, these moments are deterministic functions of the parameters.
For calculating these moments, modern distributed computational tools, e.g., graphical processing units (GPU), are available.
Given $\bm{\theta}^{*}$, each of the first and second order moments are obtained separately without passing information.
Hence, this computation would be reduced to $\mathcal{O}(N_{B})$, and need not to be a bottleneck.

Secondly, we need to generate $N_{imp}$ samples from the importance density, so we need to choose $N_{imp}$.
\cite{Doucetetal(15)} suggest $N_{imp}$ should be decided so that the standard deviation of the loglikelihood is around 1 under the assumption that the distribution of additive noise for loglikelihood estimator is Gaussian with variance inversely proportional to the number of samples and independent of the parameter value at which it is evaluated and when the MH algorithm using the exact likelihood is efficient.
In practice, checking of this assumption is uncertain.
Since the AMP algorithm does not require large $M$, $N_{imp}$ need not to be large.
When $M$ is relatively small and a large number of points are observed, the Laplace approximation for the posterior distribution of $\log \bm{\delta}$ would be close to the Gaussian distribution.
Then, the importance density is close to the posterior density of $\log \bm{\delta}$ so that $N_{imp}$ is not required to be large.

Finally, Cholesky decomposition and inversion of $M\times M$ matrices is necessary for generating samples from and evaluating the importance density via Laplace approximation.
Due to the log concavity of the log of the Poisson likelihood given intensity, the gradient descent algorithm is implemented for searching for a mode.  Then, we calculate the inverse and Cholesky decomposition of the covariance matrices. Hence, the computational cost for this step is $\mathcal{O}(M^3)$.

\subsection{Multivariate Extension}
\label{sec:Extensions}
The literature for multivariate spatial point processes is mainly restricted to the bivariate case (e.g., \cite{DiggleMilne(83)}, \cite{Allardetal(01)}, \cite{BrixMoller(01)}, \cite{Diggle(03)}, \cite{Liangetal(09)} and \cite{Picardetal(09)}).
\cite{Molleretal(98)} suggests a multivariate extension of LGCP.
\cite{BrixMoller(01)} consider a common latent process driving a conditionally independent process specification for a bivariate LGCP.
\cite{Waagepetersenetal(16)} propose a factor-like specification for a common latent process, suggesting a promising direction for higher dimensional LGCP.
The estimation strategy is based on minimum contrast estimation with respect to the pair correlation function for the mLGCP.
Due to computational cost, we find no literature on Bayesian inference for the mLGCP except for the bivariate case.

For the multivariate extension, specification of a cross covariance function (see, \cite{GentonKleiber(15)}) for the mLGCP is required.
Let $L$ is the dimension of a point pattern over the same domain, i.e., we have $L$ point patterns over the same domain.  The simplest cross covariance function is a separable form, $C_{\ell \ell^{'}}(\bm{u},\bm{v})=\rho(\bm{u},\bm{v})\gamma_{\ell \ell^{'}}, \quad \bm{u}, \bm{v}\in \mathcal{D}$ for all $\ell,\ell^{'}=1,\ldots, L$, where $\rho(\bm{u}, \bm{v})$ is a valid, non-stationary or stationary correlation function and $\gamma_{\ell \ell^{'}}=cov(Z_{\ell},Z_{\ell^{'}})$ is the nonspatial covariance between variables $\ell$ and $\ell^{'}$.
An alternative choice is the linear model of coregionalization (LMC, \cite{SchmidtGelfand(03)} and \cite{BanerjeeCarlinGelfand(14)}).
It represents a multivariate random field as a linear combination of $H\le L$ independent univariate random fields. The resulting cross-covariate functions takes the form $C_{\ell \ell^{'}}(\bm{u},\bm{v})=\sum_{h=1}^{H}\rho_{h}(\bm{u},\bm{v})\gamma_{\ell h}\gamma_{\ell^{'}h}, \quad \bm{u}, \bm{v}\in \mathcal{D}$.
Then, we can define the $\ell$-th Gaussian components inside the intensity function at location $\bm{s}$ as $z_{\ell}(\bm{s})=\sum_{h=1}^{H}\gamma_{\ell h}\nu_{h}(\bm{s})$ for $\ell=1,\ldots, L$ where $\nu_{1},\ldots,\nu_{H}$ is mean 0 and variance 1 GP with spatial correlation $\rho_{h}(\cdot)$.
The cross pair correlation function between the $\ell$-th and $\ell^{'}$-th components is $g_{\bm{\theta}}^{\ell\ell^{'}}(\bm{u},\bm{v})=\exp\biggl(\sum_{h=1}^{H}\rho_{h}(\bm{u},\bm{v})\gamma_{\ell h}\gamma_{\ell^{'}h}\biggl)$.
Furthermore, $\beta_{\bm{\theta}, mm'}^{\ell \ell^{'}}=Cov[T_{m}^{\ell}, T_{m'}^{\ell^{'}}|\bm{\theta}]$, the cross covariance for the counts of the $\ell$-th and $\ell^{'}$-th components on $B_{m}$ and $B_{m'}$ is
\begin{align}
\beta_{\bm{\theta}, mm'}^{\ell \ell^{'}}&=\bm{1}(\ell=\ell^{'})\int_{B_{m}\cap B_{m'}}\lambda_{\bm{\theta}}^{\ell}(\bm{u})d\bm{u}+\int_{B_{m}}\int_{B_{m'}}\lambda_{\bm{\theta}}^{\ell}(\bm{u})\lambda_{\bm{\theta}}^{\ell^{'}}(\bm{v})\{g_{\bm{\theta}}^{\ell \ell^{'}}(\bm{u},\bm{v})-1\}d\bm{u}d\bm{v},
\end{align}
where $\lambda_{\theta}^{\ell}(\cdot)$ and $\lambda_{\theta}^{\ell^{'}}(\cdot)$ are the intensities of the $\ell$-th and $\ell^{'}$-th components.
Importantly, the number of parameters will grow large with a mLGCP, making the INLA approach difficult to implement.
The computational cost required for multivariate extension of our algorithm is $\mathcal{O}((M\times L)^3)$.
On the other hand, if we assume an independent LGCP for each component, the computational cost still remains $\mathcal{O}(M^3)$.

\section{Simulation Studies}
\label{sec:Simulation}

In this section, we offer two simulation examples.  The first one is a univariate LGCP with different decay (or range) parameters for the GP driving the intensity surface to investigate the influence of some user specific factors and compare its performance with MCMC based inference. The second is a three dimensional LGCP which is more practically motivated and whose settings are closer to the real data application introduced later.
The algorithms in this and the next section are run on an Intel(R) Xeon(R) Processor X5675 (3.07GHz) with 12 Gbytes of memory.
\subsection{A univariate LGCP}
\label{sec:Sim1}
\begin{figure}[htbp]
  \begin{center}
   \includegraphics[width=10cm]{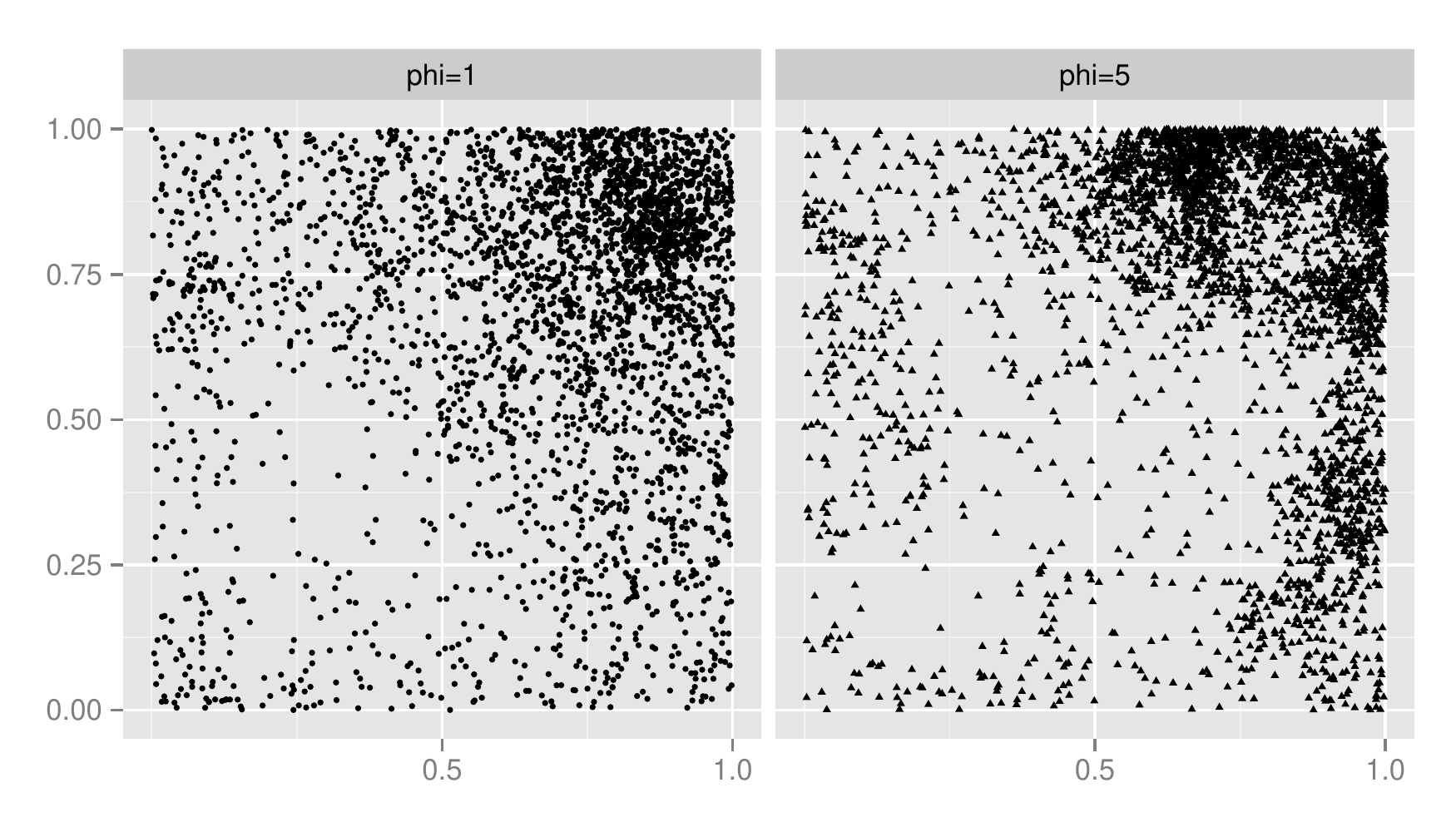}
  \end{center}
  \caption{The plot of univariate LGCP: $\phi=1$ (left) and $\phi=5$ (right)}
  \label{fig:Plot_Sim1}
\end{figure}

In this first example, we consider the univariate LGCP with different smoothness level of intensity surface and investigate the influence of the choice of $M$ and $N_{B}$ for the AMP approach.
The model we assume is
\begin{align}
\lambda(\bm{s})=\exp(\beta_{0}+\beta_{1}|s_{x}-0.3|+\beta_{2}|s_{y}-0.3|+z(\bm{s})), \quad \bm{z}(\mathcal{S})\sim \mathcal{N}(\bm{0}, \mathbf{C}_{\zeta}(\mathcal{S},\mathcal{S}))
\end{align}
where $\bm{s}=(s_{x},s_{y})$, $\mathbf{C}_{\bm{\zeta}}(\mathcal{S},\mathcal{S})=(C_{\bm{\zeta}}(\bm{s}_{i}, \bm{s}_{j}))_{i,j=1,\ldots,n}$, $C_{\bm{\zeta}}(\bm{s}_{i}, \bm{s}_{j})=\sigma^2\exp(-\phi\|\bm{s}_{i}-\bm{s}_{j}\|)$ and $\bm{\zeta}=(\sigma^2,\phi)$.
The study region/domain is defined as $\mathcal{D}=[0,1]\times [0,1]$.
The true parameter values are $(\beta_{0}, \beta_{1}, \beta_{2}, \sigma^2)=(6, 3, 3, 1)$.
We set the decay parameter at: (1) $\phi=1$ (smooth, i.e., slow decay) and (2) $\phi=5$ (rough, i.e., rapid decay).
Figure \ref{fig:Plot_Sim1} shows a realization of a univariate LGCP for each of the $\phi$s.

The numbers of simulated points for the cases are (1) $n=3,185$ and (2) $n=3,400$, respectively.
As our prior choice, we assume $\beta_{0},\beta_{1}, \beta_{2}\sim \mathcal{N}(0, 100)$ and a flat prior with sufficiently large support for $\phi$.
For the AMP approach, we discard first $i_{0}=1,000$ samples as the burn-in period  and preserve subsequent $I=5,000$ samples as posterior samples.
As for the choice of $N_{imp}$, we set $N_{imp}=1,000$. We implement adaptive random walk MH, algorithm 4 in \cite{AndrieuThoms(08)} for updating $\bm{\theta}$.
As benchmark cases, we implement an elliptical slice sampling algorithm and MMALA with a grid approximated likelihood.
Detailed fitting for both algorithms is explained in the Appendix.
We set $K=2,500$.
For elliptical slice sampling and MMALA, we discard the first $i_{0}=10,000$ samples as burn-in and retain the subsequent $I=100,000$ samples as posterior draws.

We consider different settings for $(M, N_{B})$: (i) (400, 9), (ii) (400, 1) (iii) (100, 36) and (iv)  (100, 4). For (i) and (iii), $(M, N_{B})$ are chosen so that $M\times N_{B}=3,600$. For (ii) and (iv), $(M, N_{B})$ are selected so that $M\times N_{B}=400$.
Case (i) provides a setting where $M$ is relatively large and $N_{B}$ is also taken large enough for accurate moments evaluation.
Case (ii) provides a setting where $M$ is relatively large and $N_{B}$ is taken too small for accurate moments evaluation.
Similarly, case (iii) yields relatively small grids and fine moments evaluation while case (iv) offers a situation where both approximation are coarse.

Tables \ref{tab:ResultPhi1} and \ref{tab:ResultPhi5} show the estimation results.
For both datasets, the AMP approach shows much smaller values of the inefficiency factor (IF) than elliptical slice sampling and MMALA.
The inefficiency factor is the ratio of the numerical variance of the estimate
from the MCMC samples relative to that from hypothetical uncorrelated samples, and is
defined as $1+2\sum_{t=1}^{\infty}\rho_{t}$ where $\rho_{t}$ is the sample autocorrelation at lag $t$.
It suggests the relative number of correlated draws necessary to attain the same variance of the posterior sample mean from the uncorrelated draws (e.g., \cite{Chib(01)}).
The results suggest that the AMP approach is more efficient than elliptical slice sampling and MMALA.

\begin{table}[t]
\caption{Estimation results for simulated point patterns: $\phi=1$}
\centering
\scalebox{0.9}[0.8]{
\begin{tabular}{lcccccccccc}
\hline
\hline
  & True & Mean & Stdev &  $95\%$ Int & IF &  Mean & Stdev &  $95\%$ Int & IF  \\
\hline
& & \multicolumn{2}{c}{Elliptical}    & && \multicolumn{2}{c}{MMALA} &&\\
$\beta_{0}$ & 6 & 5.974 & 0.548 & [4.788, 6.800] & 6382 & 6.232 & 0.474 & [5.370, 7.176] & 6087 \\
$\beta_{1}$ & 3 & 2.490 & 0.477 & [1.682, 3.419] & 3830 & 2.404 & 0.677 & [0.827, 3.520] & 4950 \\
$\beta_{2}$ & 3 & 3.076 & 0.876 & [1.480, 4.903] & 5963 & 2.527 & 0.764 & [1.164, 3.790] & 5810 \\
$\sigma^{2}$ & 1 & 0.796 & 0.654 & [0.236, 2.718] & 2175 & 0.630 & 0.561 & [0.204, 2.413] & 3139 \\
$\phi$ & 1 & 2.319 & 1.408 & [0.422, 5.458] & 2846 & 2.991 & 1.719 & [0.415, 6.755] & 3033 \\
$\sigma^2\phi$ & 1 & 1.203 & 0.240 & [0.807, 1.727] & 1733 & 1.214 & 0.277 & [0.763, 1.871] & 1133 \\
\hline
& & \multicolumn{2}{c}{AMP (i)}   & && \multicolumn{2}{c}{AMP (ii)} &&\\
$\beta_{0}$ & 6 & 6.721 & 0.973 & [5.034, 9.098] & 18 & 6.778 & 1.059 & [5.065, 9.291] & 19 \\
$\beta_{1}$ & 3 & 1.887 & 0.913 & [0.041, 3.691] & 11 & 1.946 & 0.873 & [0.237, 3.631] & 13 \\
$\beta_{2}$ & 3 & 2.230 & 0.902 & [0.303, 3.993] & 18 & 2.328 & 0.921 & [0.403, 3.978] & 15 \\
$\sigma^{2}$ & 1 & 1.075 & 0.828 & [0.257, 3.419] & 20 & 1.165 & 0.910 & [0.241, 3.554] & 18 \\
$\phi$ & 1 & 1.934 & 1.484 & [0.347, 5.829] & 32 & 1.402 & 1.088 & [0.271, 4.487] & 28 \\
$\sigma^2\phi$ & 1 & 1.252 & 0.301 & [0.770, 1.956] & 20 & 0.968 & 0.205 & [0.620, 1.391] & 19 \\
\hline
& & \multicolumn{2}{c}{AMP (iii)}  & && \multicolumn{2}{c}{AMP (iv)} &&\\
$\beta_{0}$ & 6 & 6.505 & 0.905 & [4.948, 8.632] & 24 & 6.582 & 0.965 & [4.677, 8.784] & 39 \\
$\beta_{1}$ & 3 & 2.110 & 0.864 & [0.334, 3.768] & 18 & 2.061 & 0.887 & [0.397, 3.917] & 13 \\
$\beta_{2}$ & 3 & 2.261 & 0.843 & [0.648, 3.910] & 28 & 2.351 & 0.869 & [0.576, 4.178] & 14 \\
$\sigma^{2}$ & 1 & 1.002 & 0.827 & [0.215, 3.406] & 20 & 1.038 & 0.844 & [0.222, 3.304] & 17 \\
$\phi$ & 1 & 1.899 & 1.606 & [0.285, 6.430] & 28 & 1.899 & 1.606 & [0.285, 6.430] & 28 \\
$\sigma^2\phi$ & 1 & 1.110 & 0.330 & [0.642, 1.999] & 20 & 0.951 & 0.224 & [0.565, 1.434] & 29 \\
\hline
\hline
\end{tabular}
}
\label{tab:ResultPhi1}
\end{table}

Across the different $(M, N_{B})$ choices, the estimation results for the AMP with $(M, N_{B})=(400, 9)$ are closer to those of elliptical slice sampling and MMALA than other AMP fits.
The posterior density of $\sigma^2\phi$ for AMP with $(M, N_{B})=(400, 1)$ shows some bias for the rough surface case.
Although AMP with $(M, N_{B})=(100, 36)$ and $(M, N_{B})=(100, 1)$ recovers the true values, the posterior distributions are flatter, especially for the rough surface cases, than with $(M, N_{B})=(400, 9)$.
Although elliptical slice sampling and MMALA with $K=2,500$ fails to recover $\sigma^2\phi$ for the case of $\phi=5$, AMP with the exception of $(M, N_{B})=(400, 1)$ recover the all true values. This suggests that $K=2,500$ is not enough for exact posterior sampling for this case.
Comparing the different $N_{B}$, small $N_{B}$ reduce the posterior variance of $\sigma^2\phi$. Especially, for the rough surface case, although AMP with $(M, N_{B})=(400, 1)$ shows the shorter credible interval (CI) for $\sigma^2\phi$ than with $(M, N_{B})=(400, 9)$, this CI doesn't include the true value. Likewise, for the $M=100$ cases, although AMP with $(M, N_{B})=(100, 4)$ recovers the true parameter value, this reflects its larger posterior variance than $M=400$ cases. In AMP approach, $M$ controls the posterior variance because $M$ can be considerd as "number of responses" and $N_{B}$ controls the biases for the posterior estimator and its variance because insufficient $N_{B}$ causes the biases in the moments calculations.

\begin{table}[t]
\caption{Estimation results for simulated point patterns: $\phi=5$. Bold denotes that $95\%$ credible interval does not include the true value.}
\centering
\scalebox{0.9}[0.8]{
\begin{tabular}{lcccccccccc}
\hline
\hline
  & True & Mean & Stdev &  $95\%$ Int & IF &  Mean & Stdev &  $95\%$ Int & IF  \\
\hline
& & \multicolumn{2}{c}{Elliptical}  & & &\multicolumn{2}{c}{MMALA} &&\\
$\beta_{0}$ & 6 & 5.591 & 0.403 & [4.789, 6.328] & 4013 & 5.706 & 0.844 & [4.286, 7.214] & 6777 \\
$\beta_{1}$ & 3 & 3.637 & 0.512 & [2.693, 4.475] & 3491 & 3.916 & 1.059 & [1.540, 5.549] & 6331 \\
$\beta_{2}$ & 3 & 3.174 & 0.683 & [1.953, 4.743] & 4693 & 2.694 & 0.956 & [1.024, 4.207] & 6335 \\
$\sigma^{2}$ & 1 & 0.814 & 0.323 & [0.455, 1.724] & 2734 & 0.945 & 0.467 & [0.432, 2.281] & 3455 \\
$\phi$ & 5 & 5.087 & 1.896 & [1.897, 8.943] & 3260 & 4.619 & 2.015 & [1.346, 9.875] & 4165 \\
$\sigma^2\phi$ & 5 & 3.641 & 0.538 & \bf{[2.712, 4.863]} & 2392 & 3.622 & 0.515 & \bf{[2.757, 4.795]} & 1096 \\
\hline
 & & \multicolumn{2}{c}{AMP (i)}  & & &\multicolumn{2}{c}{AMP (ii)} &&\\
$\beta_{0}$ & 6 & 6.010 & 0.732 & [4.653, 7.690] & 11 & 6.337 & 1.168 & [4.426, 9.155] & 36 \\
$\beta_{1}$ & 3 & 3.576 & 1.073 & [1.351, 5.716] & 14 & 3.440 & 1.247 & [1.116, 5.926] & 17 \\
$\beta_{2}$ & 3 & 2.479 & 1.140 & [0.025, 4.512] & 14 & 2.203 & 1.244 & [-0.473, 4.632] & 21 \\
$\sigma^{2}$ & 1 & 1.037 & 0.514 & [0.499, 2.517] & 22 & 1.319 & 0.836 & [0.486, 3.535] & 41 \\
$\phi$ & 5 & 5.166 & 2.197 & [1.429, 10.13] & 17 & 2.736 & 1.467 & [0.663, 6.556] & 33 \\
$\sigma^2\phi$ & 5 & 4.504 & 0.847 & [2.994, 6.304] & 15 & 2.683 & 0.432 & \bf{[1.934, 3.640]} & 21 \\
\hline
 & & \multicolumn{2}{c}{AMP (iii)}  & & &\multicolumn{2}{c}{AMP (iv)} &&\\
$\beta_{0}$ & 6 & 5.899 & 0.800 & [4.628, 7.790] & 14 & 5.968 & 0.919 & [4.514, 8.418] & 50 \\
$\beta_{1}$ & 3 & 3.710 & 1.143 & [1.231, 5.903] & 14 & 3.852 & 1.141 & [1.597, 6.071] & 17 \\
$\beta_{2}$ & 3 & 2.725 & 1.145 & [0.144, 4.810] & 36 & 2.671 & 1.228 & [-0.118, 4.862] & 12 \\
$\sigma^{2}$ & 1 & 1.108 & 0.501 & [0.566, 2.546] & 49 & 1.151 & 0.634 & [0.498, 2.865] & 24 \\
$\phi$ & 5 & 5.297 & 2.448 & [1.475, 10.53] & 36 & 4.219 & 2.126 & [1.167, 9.230] & 44 \\
$\sigma^2\phi$ & 5 & 4.996 & 1.394 & [3.005, 8.582] & 21 & 3.914 & 0.997 & [2.380, 6.369] & 45 \\
\hline
\hline
\end{tabular}
}
\label{tab:ResultPhi5}
\end{table}

Wider credible intervals are also reported in \cite{FilipponeGirolami(14)}.
The posterior distributions of the parameters with MCMC based algorithms including elliptical slice sampling are sharply peaked and highly dependent on the sampled GP (\cite{MurrayAdams(10)} and \cite{FilipponeGirolami(14)}).
Since the estimation results of MMALA show high IFs which are close to elliptical slice sampling, we may not have reached convergence to the stationary distribution for both MMALA and elliptical slice sampling.

Overall, the estimation results suggest that even if $M$ is of moderate dimension, the AMP approach recovers the posterior distribution as long as $N_{B}$ is fine enough. Also, the AMP approach with sufficiently large $M$ and $N_{B}$ captures the true values even in the rough surface case where finer resolution (larger $K$) over the study region is required.



\begin{table}[htbp]
\caption{Computational Efficiency for the simulated data}
\label{tab:Sim1_Eff}
\centering
\scalebox{0.9}[0.8]{
\begin{tabular}{lcccccc}
\hline
\hline
 Method   & Time & IF         & (Time/$I$) & (Time/$I$)$\times$IF \\
              &   (sec)  & (min, max) & (sec) & (min, max) \\
\hline
$\phi=1$ &  &  & & \\
Elliptical  & 324446 & (1133, 6087) & 2.94 & (3341, 17953) \\
MMALA  & 347672 & (1733, 6382) & 3.16 & (5477, 20171) \\
AMP (i) & 15748 & (11, 32) & 3.14 & (34.5, 100) \\
AMP (ii) & 11780 & (13, 28) & 2.35 & (30.6, 65.9) \\
AMP (iii) & 4360 & (18, 28) & 0.87 & (15.6, 24.4) \\
AMP (iv) & 815 & (13, 39) & 0.16 & (2.1, 6.3) \\
\hline
$\phi=5$ &  &  & & \\
Elliptical  & 329205 & (2392, 4693) & 2.99 & (7158, 14045) \\
MMALA  & 347933 & (1096, 6777) & 3.16 & (3466, 21435) \\
AMP (i) & 16086 & (11, 22) & 3.21 & (35.3, 70.6) \\
AMP (ii) & 11221 & (17, 41) & 2.24 & (38.1, 92.0) \\
AMP (iii) & 4717 & (14, 49) & 0.94 & (13.2, 46.2) \\
AMP (iv) & 855 & (12, 50) & 0.17 & (2.0, 8.5) \\
\hline
\hline
\end{tabular}
}
\end{table}

Comparing the efficiency of algorithms is difficult. Here, we focus on comparing the computational efficiency of elliptical slice sampling, MMALA and AMP approach.
We use the "(computational time for one iteration (Time/$I$)) $\times$ (Inefficiency Factor)", which corresponds to the approximate time for generating one independent sample from the stationary distribution.
The AMP approach is implemented without parallelization of calculation of moments, so the results for the AMP approach are conservative.
Table $\ref{tab:Sim1_Eff}$ shows the computational efficiency of each approach.
"Time" means the computational time (seconds) for generating $I$ posterior samples (we ignore the burn-in period here).
We show the minimum and maximum of inefficiency factor for parameters in Table $\ref{tab:ResultPhi1}$ and $\ref{tab:ResultPhi5}$.
Overall, AMP approaches are far faster than elliptical slice sampling and MMALA approaches. For AMP approaches, even with $(M=400, N_{B}=9)$, at most 100 seconds is required for generating one independent approximate posterior samples. Finally, the IF factors may be helpful but, due to dependence on how the actual computing environments were configured for the different approaches, it is difficult to attempt to meaningfully compare run times.
As for the comparison of the performance of INLA and MCMC based approach for a spatial LGCP, \cite{TaylorDiggle(14)} compare the predictive accuracy of MCMC and INLA for a spatial LGCP with fixed parameters values.
In this setting, they found that MCMC provides more accurate estimates of predictive probabilities than INLA approach with 100,000 iterations.
However, only comparing the estimation accuracy may not be enough because it ignores the rapid computation that INLA provides.

\subsection{A multivariate LGCP}
\label{sec:Sim2}

\begin{figure}[htbp]
  \begin{center}
   \includegraphics[width=15cm]{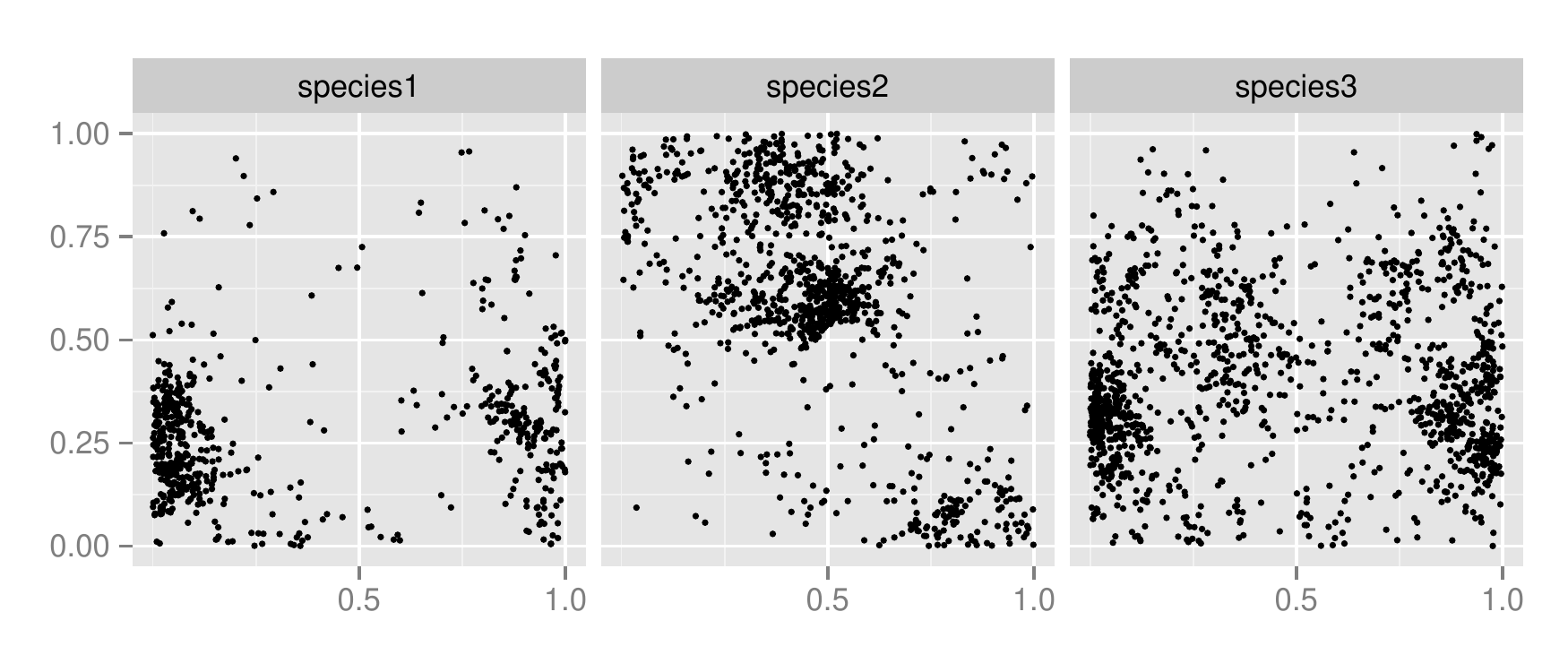}
  \end{center}
  \caption{The plot of the simulated three dimensional LGCP}
  \label{fig:Plot_Sim2}
\end{figure}

As a second example, we consider a mLGCP (\cite{Molleretal(98)}), a three dimensional LGCP.
This model introduces larger hyperparameter dimensionality than the univariate LGCP.
MCMC based inferences for this model are difficult due to their huge computational costs. So, in this example and the real data application in the next section, we focus only on implementation of AMP approach without elliptical slice sampling and MMALA.
In fact, we implement a model similar to that for the tree species dataset in the next section and the parameter settings are based on the real data application described in section \ref{sec:Real}. The simulation model is defined as
\begin{align}
\begin{aligned}
\bm{\lambda}(\bm{s})&=\exp\biggl(\bm{1}_{3}\beta_{0}+\mathbf{\Gamma}\bm{\nu}(\bm{s})\biggl), \\
\bm{\nu}_{i}(\mathcal{S}) &\sim \mathcal{N}(\bm{0}, \mathbf{C}_{\phi_{i}}(\mathcal{S},\mathcal{S}))
\end{aligned}
\quad
\begin{aligned}
\mathbf{\Gamma}=\begin{pmatrix}
                2 &  0 & 0 \\
                -1 &  1 & 0 \\
                   1 &  0 & 1
                \end{pmatrix}, \quad \bm{\nu}(\bm{s})=\begin{pmatrix}
                                                        \nu_{1}(\bm{s}) \\
                                                        \nu_{2}(\bm{s}) \\
                                                        \nu_{3}(\bm{s}) \\
                                                      \end{pmatrix},
\end{aligned}
\end{align}
where $\bm{\nu}_{i}$ is independent zero mean and unit variance GP with isotropic exponential correlation function having decay parameters $\phi_{i}$, respectively.
The correlation between the first and corresponding rows is introduced from the latent process $\nu_{1}$ and individual behavior is captured by $\nu_{2}$ and $\nu_{3}$.
When $\gamma_{i1}$ and $\gamma_{j1}$ have different signs, the $i$-th and $j$-th point patterns show negative dependence, i.e, the cross pair correlation function is smaller than 1 (e.g., \cite{Molleretal(98)} and \cite{Illianetal(08)}).
The number of simulated points for each component is $N_{1}=639$, $N_{2}=1,050$ and $N_{3}=1,283$.
Figure \ref{fig:Plot_Sim2} shows the three jointly simulated point patterns.
The negative dependence between first and second point pattern is observed, which is induced by the $\gamma_{21}=-1$.
We set $M=144$, $N_{B}=25$ and $N_{imp}=1,000$.

\begin{figure}[htbp]
\def\@captype{table}
\begin{minipage}{.48\textwidth}
\tblcaption{Estimation results for the simulated three dimensional LGCP}
\label{tab:ResultSim2}
\centering
\scalebox{0.9}[0.8]{
\begin{tabular}{lccccccc}
\hline
\hline
       & True & Mean & Stdev & $95\%$ Int & IF \\
\hline
\multicolumn{4}{c}{AMP $(M, N_{B})=(144,25)$} & \\
$\beta_{0}$ & 7 & 7.200 & 0.491 & [6.353, 8.295] & 24 \\
$\phi_{1}$ & 3 & 3.056 & 1.001 & [1.582, 5.497] & 21 \\
$\phi_{2}$ & 5 & 3.540 & 1.793 & [1.118, 7.576] & 24 \\
$\phi_{3}$ & 5 & 3.659 & 1.978 & [1.105, 8.518] & 26 \\
$\gamma_{11}$ & 2 & 1.933 & 0.231 & [1.519, 2.413] & 11 \\
$\gamma_{21}$ & -1 & -1.130 & 0.246 & [-1.676, -0.688] & 19 \\
$\gamma_{31}$ & 1 & 1.192 & 0.218 & [0.795, 1.629] & 42 \\
$\gamma_{22}$ & 1 & 1.335 & 0.280 & [0.929, 2.074] & 8 \\
$\gamma_{33}$ & 1 & 1.109 & 0.247 & [0.752, 1.710] & 27 \\
\hline
\hline
\end{tabular}
}
\end{minipage}
\hfill
\begin{minipage}{.48\textwidth}
\figcaption{The plot of demeaned surface of \texttt{elevation}}
  \begin{center}
   \includegraphics[width=7cm]{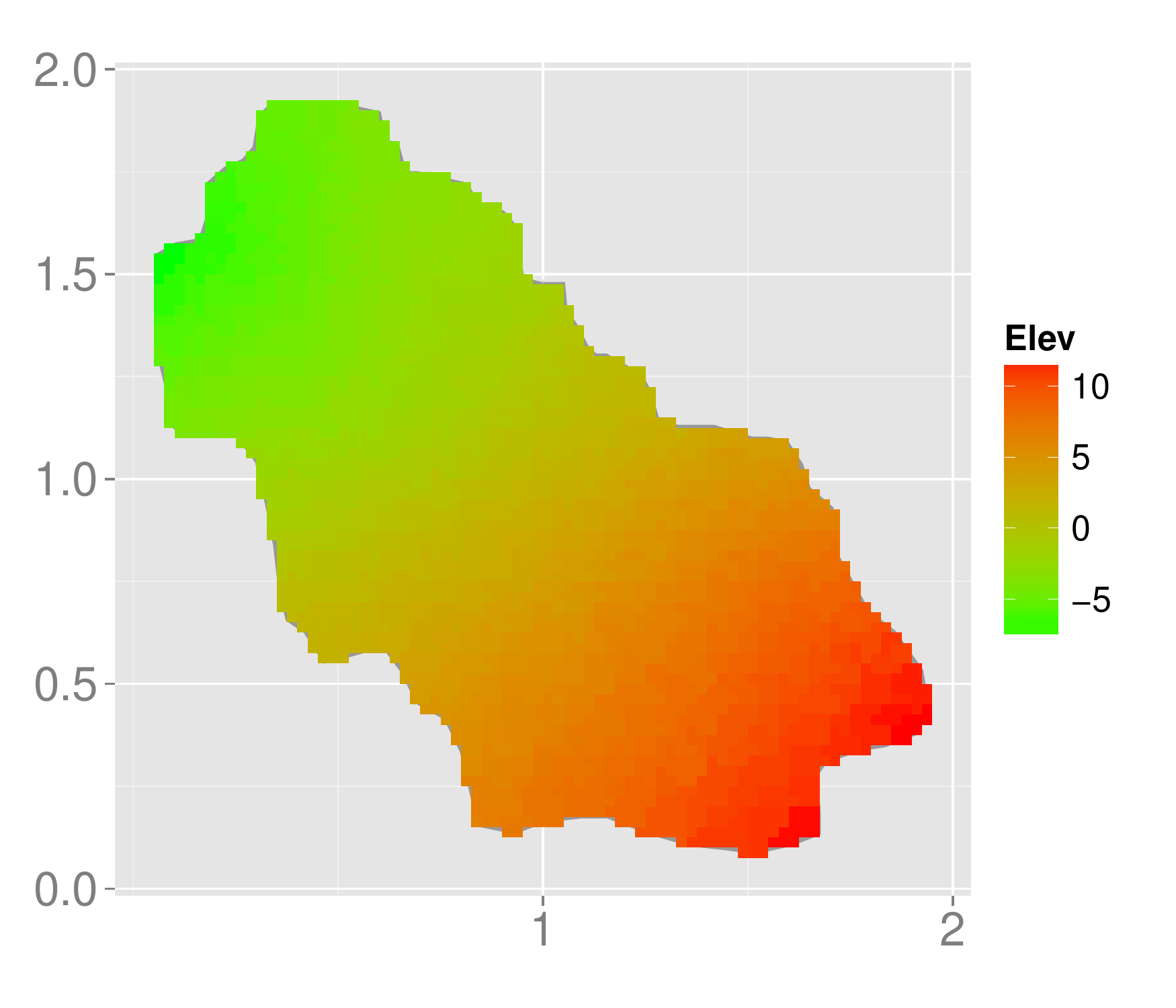}
  \end{center}
  \label{fig:Plot_Real_Elev}
\end{minipage}
\end{figure}

We assume $\beta_{0} \sim \mathcal{N}(0, 100)$, $\gamma_{ij}^2\sim \mathcal{G}(2,1)$ and a flat prior with sufficiently large support for $\phi_{i}$.
Due to the lack of identifiability for the variance parameter and the decay parameter in the GP (\cite{Zhang(04)}), we put a relatively informative prior on the components of $\mathbf{\Gamma}$, $(\gamma_{ij})_{i,j=1,2,3}$.
The burn in period is $i_{0}=1,000$ and $I=5,000$ samples are preserved as posterior samples.
As in the univariate LGCP, we implement adaptive random walk MH, algorithm 4 in \cite{AndrieuThoms(08)} for parameter estimation.
The computational cost of the AMP approach is $\mathcal{O}((3M)^3)$ for implementing Laplace approximation.
Table \ref{tab:ResultSim2} shows the estimation results for the data. Our approach recovers parameter values well with small inefficiency factors.
The posterior variances for $\phi_{2}$ and $\phi_{3}$ are larger than for $\phi_{1}$.

\section{Application}
\label{sec:Real}
\begin{figure}[htbp]
  \begin{center}
   \includegraphics[width=15cm]{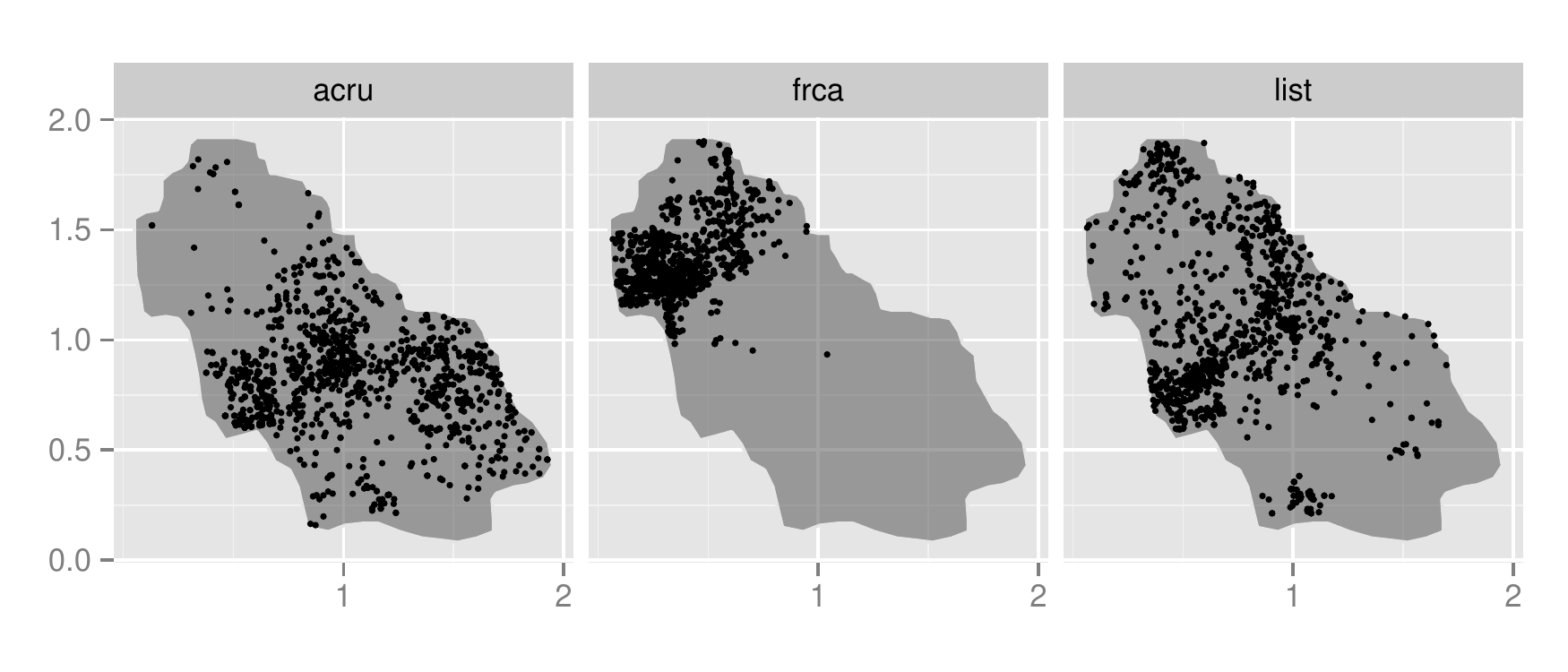}
  \end{center}
  \caption{The plot of point patterns of three species in Duke forest: 1. Red Maple (left), 2. Carolina Buckthorn (middle) and 3. Sweetgum (right)}
  \label{fig:Plot_Real}
\end{figure}

In this section, we implement our approach for a real dataset.
Our overall dataset consists of tree locations in Duke Forest, comprising 63 species, a total of 14,992 individuals with recorded location.
We select 3 prevalent species yielding relatively large number of points: (i). {\it acer rubrum} (\texttt{acru}, Red Maple), (ii). {\it frangula caroliniana} (\texttt{frca}, Carolina Buckthorn) and (iii). {\it liquidambar styraciflua} (\texttt{list}, Sweetgum), and embed the study region $\mathcal{D}$ onto $[0,2]\times [0,2]$.
The numbers of points in each pattern are $N_{1}=1,251$, $N_{2}=1,120$ and $N_{3}=1,067$, respectively.
Again, we consider the mLGCP as discussed in example 2 of the simulation studies.
As a covariate, we include demeaned elevation (\texttt{elevation}) for each point, its surface is in figure \ref{fig:Plot_Real_Elev}.
Figure \ref{fig:Plot_Real} shows the observed point patterns.
Negative dependence between \texttt{acru} and \texttt{frca} is expected from the figure.

We consider a three dimensional LGCP, i.e., $\bm{\lambda}(\bm{s})=\exp\biggl(\bm{1}_{3}\beta_{0}+\mathbf{X}(\bm{s})\bm{\beta}_{1}+\mathbf{\Gamma}\bm{\nu}(\bm{s})\biggl)$ where $\mathbf{X}$ is a 3$\times 3$ matrix whose $i$-th diagonal element is \texttt{elevation}$_{i}$ and for $i=1,2,3$. We assume the same specification for $\mathbf{\Gamma}$ and $\bm{\nu}(\bm{s})$ in Section \ref{sec:Sim2}.
We also estimate an independent three dimensional LGCP, i.e., $\gamma_{21}=\gamma_{31}=0$, with our AMP approach.
We discard $i_{0}=1,000$ samples as burn-in period and preserve $I=5,000$ samples as posterior samples.
For the grid selection, at first, we take an $80\times 80$ regular grid over $[0,2]\times[0,2]$,
and preserve the grid points inside $\mathcal{D}$.
The total number of grid points inside $\mathcal{D}$ is $K=3,052$.  Then, we construct larger blocks so that $M=120$, i.e., we aggregate about 25 blocks for each larger block.
$N_{B}$ is dependent on each aggregated block, i.e., $N_{B_{m}}\approx 25$ for $m=1,\ldots, M$. We set $N_{imp}=1,000$.
Again, we implement an adaptive random walk MH, algorithm 4 in \cite{AndrieuThoms(08)}.
The prior specification is $\bm{\beta}=(\beta_{0},\bm{\beta}_{1}) \sim N(\bm{0}, 100\mathbf{I}_{4})$, $\gamma_{ij}^2\sim \mathcal{G}(2,1)$ and flat priors for the $\phi_{i}$ with large support.

Table \ref{tab:Real1} shows the estimation results.
The estimated value of $\gamma_{21}$ is negative implying that \texttt{acru} and \texttt{frca} are negatively correlated as Figure \ref{fig:Plot_Real} suggested.
The inefficiency factors for the parameters with the AMP approach are small.
The result for a independent LGCP model shows smaller variance for $\phi_{2}$ and $\phi_{3}$ than that for the dependent LGCP.
The effect of \texttt{elevation} is different among species, significantly negative except for \texttt{acru}.

\begin{table}[htbp]
\caption{Estimation results for Duke forest dataset: AMP for independent mLGCP (left) and for dependent mLGCP (right)}
\centering
\scalebox{0.9}[0.8]{
\begin{tabular}{lcccccccccc}
\hline
\hline
  &  \multicolumn{4}{c}{AMP-Ind} & & \multicolumn{4}{c}{AMP-Dep}   \\
\hline
   & Mean & Stdev &  $95\%$ Int & IF & & Mean & Stdev &  $95\%$ Int & IF  \\
\hline
$\beta_{0}$ & 5.953 & 0.430 & [4.949, 6.773] & 17 &  & 5.728 & 0.472 & [4.848, 6.704] & 45 \\
$\beta_{1,1}$ & 0.085 & 0.095 & [-0.095, 0.259] & 13 &  & 0.094 & 0.116 & [-0.124, 0.325] & 33 \\
$\beta_{1,2}$ & -0.618 & 0.211 & [-1.056, -0.253] & 13 & & -0.606 & 0.166 & [-0.951, -0.290] & 20 \\
$\beta_{1,3}$ & -0.188 & 0.074 & [-0.338, -0.058] & 21 &  & -0.191 & 0.078 & [-0.354, -0.020] & 22 \\
$\phi_{1}$ & 3.156 & 1.167 & [1.190, 5.846] & 39 & & 1.798 & 0.776 & [0.740, 3.682] & 14 \\
$\phi_{2}$ & 2.594 & 0.967 & [0.972, 4.803] & 22 & & 5.335 & 3.232 & [1.535, 12.47] & 29 \\
$\phi_{3}$ & 4.159 & 1.367 & [1.869, 7.135] & 37 & & 6.202 & 2.513 & [2.143, 11.37] & 22 \\
$\gamma_{11}$ & 1.616 & 0.272 & [1.230, 2.235] & 52 & & 1.883 & 0.326 & [1.355, 2.614] & 26 \\
$\gamma_{22}$ & 2.313 & 0.324 & [1.747, 2.957] & 28 & & 1.858 & 0.355 & [1.269, 2.652] & 51 \\
$\gamma_{33}$ & 1.452 & 0.193 & [1.127, 1.890] & 36 & & 1.135 & 0.170 & [0.852, 1.567] & 57 \\
$\gamma_{21}$ &  &  &  &  & & -1.083 & 0.364 & [-1.737, -0.361] & 19 \\
$\gamma_{31}$ & &  &  & & & 1.093 & 0.277 & [0.597, 1.640] & 16 \\
\hline
\hline
\end{tabular}
}
\label{tab:Real1}
\end{table}
\section{Brief discussion}

We proposed a two stage model fitting approach for univariate and multivariate LGCPs which is based on developing an approximate marginal posterior (AMP) distribution for the model parameters.
Through the simulation studies, we demonstrated that our approach recovers the true parameter values even with relatively low dimensional grids.
Furthermore, the results suggest that the AMP approach is applicable for mLGCP models, which introduce many more parameters than the univariate case.
Comparison with the two competitive approaches - elliptical slice sampling and MMALA  - suggest improved efficiency for AMP in the univariate case.
Elliptical slice sampling and MMALA become computationally infeasible for the multivariate case.
The AMP computational scheme is similar to INLA in spirit, but INLA might not be well-suited to the multivariate case.
The AMP approach also obtains sample based joint posterior distribution $\bm{z}$, which enable us to estimate nonlinear or joint posterior probability of $\bm{z}$.
In fact, given posterior $\bm{\theta}$s, we can sample $\bm{z}|\mathcal{S}$ efficiently in a parallel computation scheme with elliptical slice sampling or MALA without the fine tuning.

\section*{Acknowledgements}
The work of the first author was supported in part by the Nakajima Foundation.
The authors thank James Clark, Jordan Siminitz for providing the Duke Forest dataset and Akihiko Nishimura, H\r{a}vard Rue for helpful comments.
The computational results were obtained by using Ox version 7.1 (\cite{Doornik(07)}).

\section*{Appendix}
\subsection*{A: MCMC based Posterior Sampling for LGCP}
In fitting the LGCP model, we have a stochastic integral of the form $\int_{\mathcal{D}}\lambda(\bm{u}|\bm{\theta}, z(\bm{u})d\bm{u}$ in the exponential of the likelihood.
We use grid cell approximation for this integral as well as for the product term in the likelihood yielding
\begin{align}
\mathcal{L}(\mathcal{S}|\bm{\theta}, \bm{z})=\exp\biggl(|D|-\sum_{k=1}^{K}\lambda(\bm{u}_{k}|\bm{\theta}, z(\bm{u}_{k}))\Delta_{k} \biggl)\prod_{k=1}^{K}\lambda(\bm{u}_{k}|\bm{\theta}, z(\bm{u}_{k}))^{n_{k}} \quad \bm{u}_{1},\ldots,\bm{u}_{K} \in \mathcal{D}
\end{align}
where $n_{k}$ is the number of events in grid $k$, $K$ is the number of grid cells, $n=\sum_{k=1}^{K}n_{k}$ is the total number of points in the point pattern, $\bm{u}_{k}$ is the centroid of the $k$-th grid cell and $\Delta_{k}$ is the area of $k$-th grid.
Although fitting this approximation is straightforward because we only require evaluation of $\lambda(\bm{u}_{k}|\bm{\theta}, z(\bm{u}_{k}))$ over the grid cells, sampling of a $K$ dimensional vector from a GP would be a computational burden. We consider two types of MCMC based sampling strategies for the GP.

One approach is the manifold Metropolis adjusted Langevin algorithm (MMALA, \cite{GirolamiCalderhead(11)}) for the LGCP.
\cite{GirolamiCalderhead(11)} implement MMALA for the simulated LGCP and show its computational efficiency for sampling of a GP relative to standard MALA with fixed $\bm{\beta}$ and $\bm{\zeta}$ at true values.
\cite{Diggleetal(13)} and \cite{Tayloretal(15)} modify the implementation of  MMALA in \cite{GirolamiCalderhead(11)} for a LGCP with sampling $\bm{\beta}$ and $\bm{\zeta}$.
We basically follow the discussion in \cite{Diggleetal(13)} and \cite{Tayloretal(15)}.
We assume
\begin{align}
\log \lambda(\bm{s}|\bm{\theta}, z(\bm{s}))&=\bm{X}(\bm{s})\bm{\beta}+z(\bm{s}), \quad \bm{z}(\mathcal{S})\sim \mathcal{N}(\bm{0}, \mathbf{C}_{\bm{\zeta}}(\mathcal{S},\mathcal{S})) \quad \bm{s}\in \mathcal{D} \\
\bm{z}&=\mathbf{L}_{\bm{\zeta}}\bm{\nu}, \quad \mathbf{L}_{\bm{\zeta}}\mathbf{L}_{\bm{\zeta}}^{'}=\mathbf{C}_{\bm{\zeta}}
\end{align}
where $\mathbf{L}_{\bm{\zeta}}$ is the Cholesky decomposition of $\mathbf{C}_{\bm{\zeta}}$, $\bm{\theta}=(\bm{\nu}, \bm{\beta},\bm{\zeta})$, $\bm{\zeta}$ is the parameter vector for the GP, $\bm{\nu}$ is a zero mean and unit variance normal vector.
\cite{Diggleetal(13)} discuss the preconditioning matrix,  $\mathbf{M}^{-1}$ (\cite{GirolamiCalderhead(11)}), to define the proposal for MMALA.
\begin{align}
q(\bm{\theta}^{*}|\bm{\theta}^{(i-1)})=\mathcal{N}(\bm{\omega}^{(i-1)}, \mathbf{\Omega}^{(i-1)})
\end{align}
where
\begin{align}
\bm{\omega}^{(i-1)}=\begin{pmatrix}
                             \bm{\nu}^{(i-1)}+\frac{\sigma_{0}^2\sigma_{\bm{\nu}}^{2}\mathbf{M}^{-1}_{\bm{\nu}}\partial \log \pi (\bm{\theta}^{(i-1)}|\mathcal{S})/\partial \bm{\nu}}{2} \\
                             \bm{\beta}^{(i-1)}+\frac{\sigma_{0}^2\sigma_{\bm{\beta}}^{2}\mathbf{M}^{-1}_{\bm{\beta}}\partial \log \pi (\bm{\theta}^{(i-1)}|\mathcal{S})/\partial \bm{\beta} }{2} \\
                            \bm{\zeta}^{(i-1)}
                            \end{pmatrix}, \quad
\mathbf{\Omega}^{(i-1)}=\sigma_{0}^2\begin{pmatrix}
                                        \sigma_{\bm{\nu}}^{2}\mathbf{M}^{-1}_{\bm{\nu}} & 0 & 0 \\
                                          0 & \sigma_{\bm{\beta}}^{2}\mathbf{M}^{-1}_{\bm{\beta}} & 0 \\
                                          0 & 0 & c\sigma_{\bm{\zeta}}^{2}\mathbf{\Sigma}_{\bm{\zeta}}
                                         \end{pmatrix}
\end{align}
They set $\sigma_{\bm{\nu}}^{2}=1.65^2/\text{dim}(\bm{\nu})^{1/3}$, $\sigma_{\bm{\beta}}^{2}=1.65^2/\text{dim}(\bm{\beta})^{1/3}$ and $\sigma_{\bm{\zeta}}^{2}=2.38^2/\text{dim}(\bm{\zeta})$.
They also set $c=0.4$ and tune $\sigma_{0}^2$ adaptively so that the acceptance rate of 0.574 is achieved. Although they consider common $\sigma_{0}^2$ and sample all parameters simultaneously, we consider, independently, $\sigma_{0}^2$ for $\bm{\nu}$ and $\sigma_{1}^2$ for $(\bm{\beta}, \bm{\zeta})$ and separately sample $\bm{\nu}$ and $(\bm{\beta}, \bm{\zeta})$ to improve mixing.
The main computational cost is calculation of $\mathbf{M}^{-1}_{\bm{\nu}}$ for high dimensional $\bm{\nu}$.
$\mathbf{M}^{-1}_{\bm{\nu}}$ and $\mathbf{M}^{-1}_{\bm{\beta}}$ are the negative inverse of the Fisher information matrix with respect to $\bm{\nu}$ and $\bm{\beta}$,
\begin{align}
\mathbf{M}_{\bm{\nu}}&=-E_{\bm{y},\bm{\nu}|\bm{\beta}, \bm{\zeta}}(\nabla_{\bm{\nu}}\nabla_{\bm{\nu}} \mathcal{L})=\mathbf{L}_{\bm{\zeta}}'\mathbf{D}\mathbf{L}_{\bm{\zeta}}+\mathbf{I} \\
\mathbf{M}_{\bm{\beta}}&=-E_{\bm{y},\bm{\nu}|\bm{\beta}, \bm{\zeta}}(\nabla_{\bm{\beta}}\nabla_{\bm{\beta}} \mathcal{L})=\mathbf{X}'\mathbf{D}\mathbf{X}+(1/\kappa_{\bm{\beta}})\mathbf{I}
\end{align}
where $\kappa_{\bm{\beta}}$ is the prior variance for $\bm{\beta}$ and $\mathbf{D}$ is the diagonal matrix whose $(k,k)$ element is $\exp(\bm{X}(\bm{s}_{k})\bm{\beta}+\sigma^2/2)\Delta_{k}$.

In practice, calculation of $\mathbf{M}^{-1}_{\bm{\nu}}$ within each MCMC iteration is computationally expensive because $\mathbf{M}^{-1}_{\bm{\nu}}$ is high dimensional and dense.
\cite{Tayloretal(15)} consider initial guesses at $\bm{\nu}$ and $\bm{\beta}$ followed by a quadratic approximation to the target.
Instead, we calculate the initial guesses of $\mathbf{M}^{-1}_{\bm{\nu}}$ and  $\mathbf{M}^{-1}_{\bm{\beta}}$ by a slightly modified strategy. Furthermore, we fix $\mathbf{M}^{-1}_{\bm{\nu}}$ but update $\mathbf{M}^{-1}_{\bm{\beta}}$ within MCMC iteration.
Due to the log-concavity of the likelihood, the maximum a posteriori (MAP) estimator is available (\cite{Molleretal(98)}).
We follow the discussion in \cite{Molleretal(98)} to search for the MAP estimators where $\mathbf{M}^{-1}_{\bm{\nu}}$ and $\mathbf{M}^{-1}_{\bm{\beta}}$ are evaluated at the initial step.
At first, since the gradient with respect to $\bm{\zeta}$ is hard to calculate, we fix initial values of  $\bm{\zeta}$ at the minimum contrast estimator based on pair correlation function for LGCP by using \texttt{spatstat} package (\cite{BaddeleyTurner(05)}).
Given the fixed $\bm{\zeta}$, the gradient decent algorithm is available as a mode search for both $\bm{\nu}$ and $\bm{\beta}$ (see, \cite{Molleretal(98)}), i.e., iterate $\bm{\nu}^{(i)}=\bm{\nu}^{(i-1)}+\delta_{\bm{\nu}} \partial \log \pi (\bm{\theta}^{(i-1)}|\mathcal{S})/\partial \bm{\nu}$ and $\bm{\beta}^{(i)}=\bm{\beta}^{(i-1)}+\delta_{\bm{\beta}} \partial \log \pi (\bm{\theta}^{(i-1)}|\mathcal{S})/\partial \bm{\beta}$ until convergence, where $\delta_{\bm{\nu}}$ and $\delta_{\bm{\beta}}$ are bandwidth parameters.
Since the algorithm diverges under large values of bandwidth parameters, we tune $\delta_{\bm{\nu}}$ and $\delta_{\bm{\beta}}$ small enough for each dataset.
Calculated initial values (MAP estimator for $(\bm{\nu},\bm{\beta})$ and minimum contrast estimator for $\bm{\zeta}$) of $(\bm{\nu},\bm{\beta},\bm{\zeta})$ are also used in the elliptical slice sampling algorithm.

In addition to MMALA for sampling the GP, we implement elliptical slice sampling (\cite{MurrayAdamsMacKay(10)} and \cite{MurrayAdams(10)}) as discussed in {\cite{LeiningerGelfand(16)}} for a spatial LGCP.
We sample $\bm{\nu}^{*}=\bm{\nu} \cos(\omega)+\bm{\eta} \sin(\omega)$ where $\bm{\eta}\sim \mathcal{N}(\bm{0}, \mathbf{I})$ and $\omega \in [0,2\pi)$ through the elliptical slice sampling algorithm (see, \cite{MurrayAdamsMacKay(10)}).
As for sampling of $(\bm{\beta},\bm{\zeta})$, we follow the same approach as in MMALA, as explained above.

\subsection*{B: Figures}

\begin{figure}[htbp]
  \begin{center}
   \includegraphics[width=13cm]{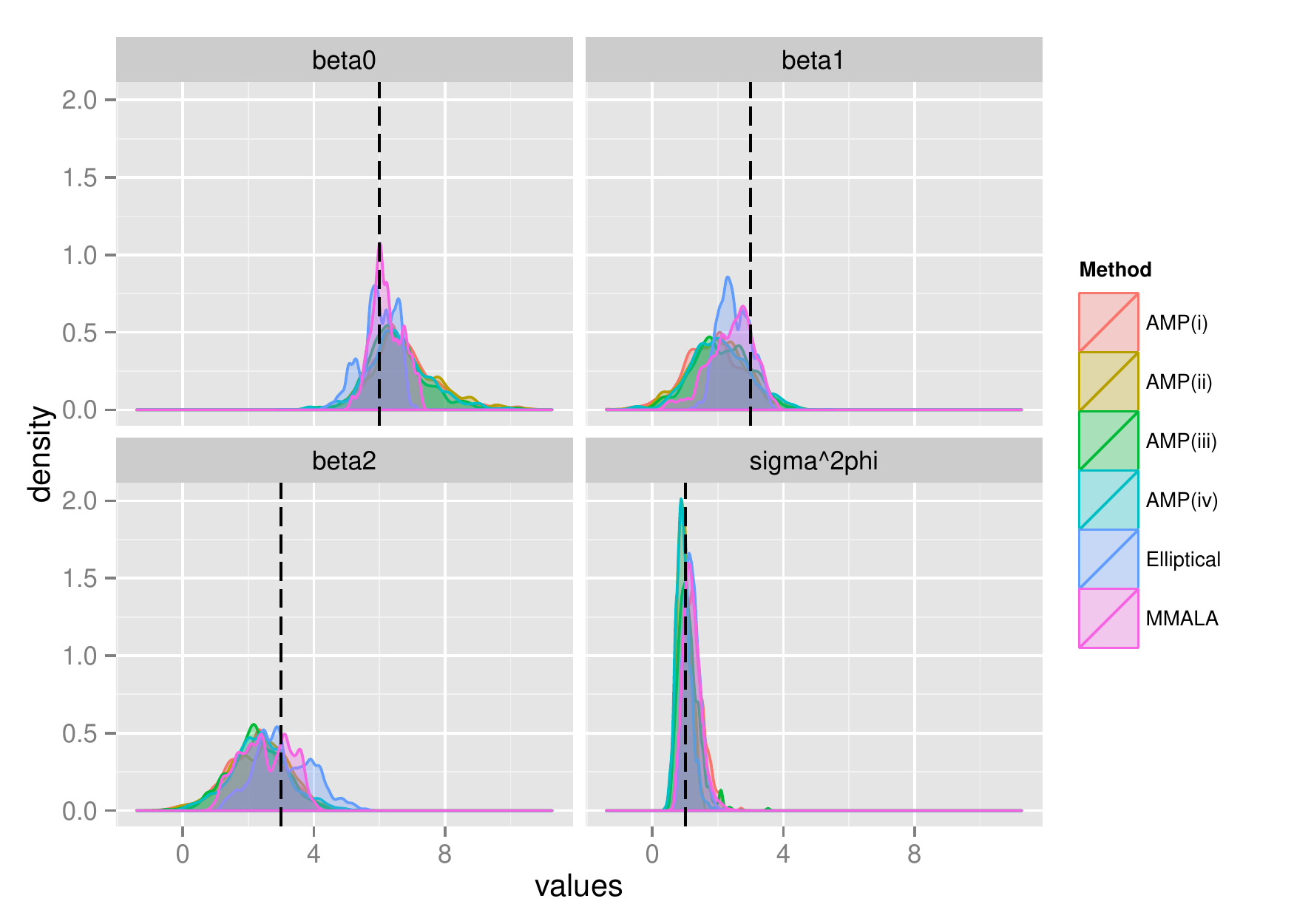}
  \end{center}
  \caption{The plot of posterior density for the univariate LGCP: $\phi=1$. Black dashed lines are true values}
\label{fig:PosDensPhi1}
\end{figure}

\begin{figure}[htbp]
  \begin{center}
   \includegraphics[width=13cm]{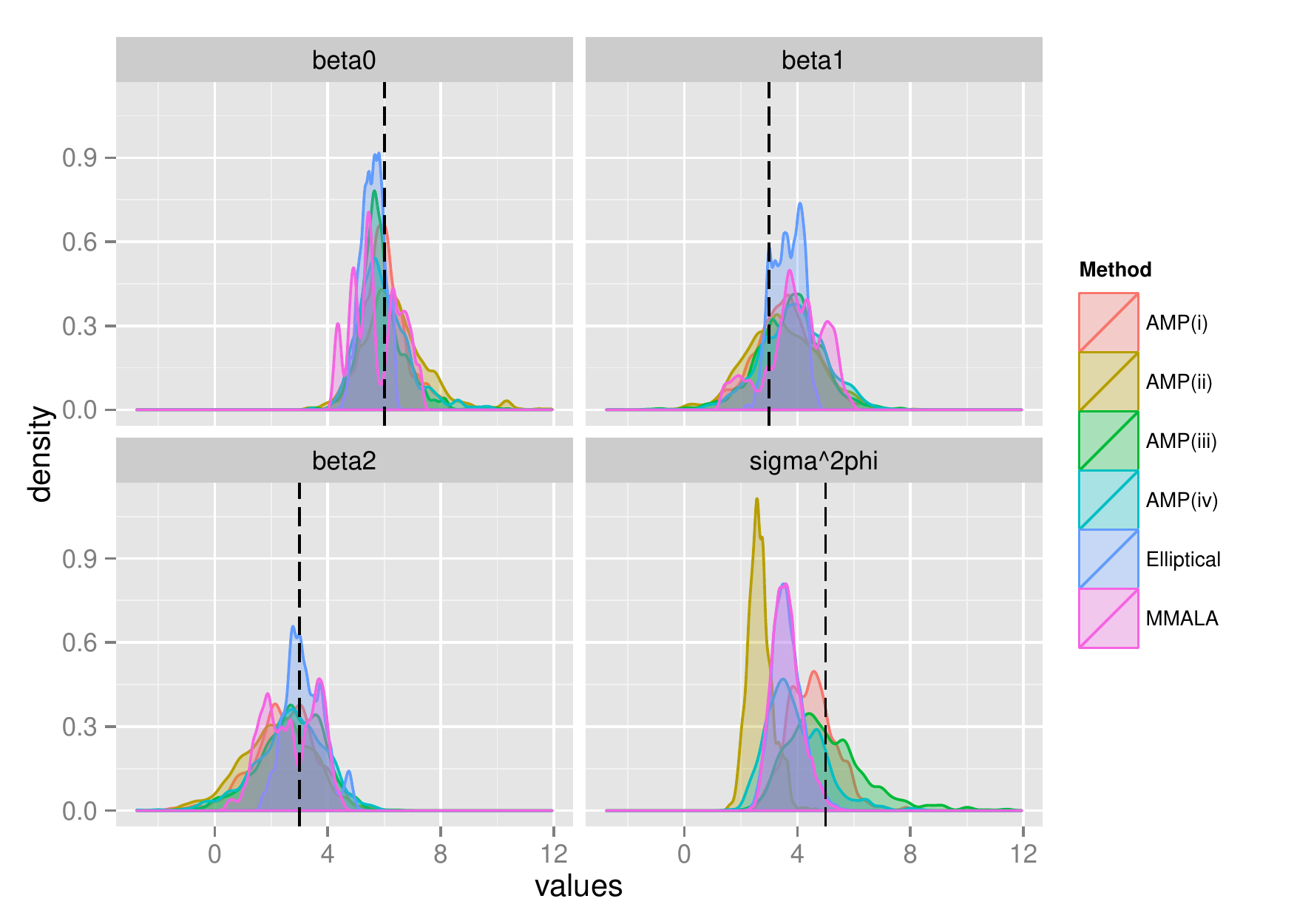}
  \end{center}
  \caption{The plot of posterior density for the univariate LGCP: $\phi=5$. Black dashed lines are true values}
\label{fig:PosDensPhi5}
\end{figure}

\bibliographystyle{chicago}
\bibliography{SP}

\end{document}